\pdfoutput=43

\newcommand{\balpha}{ \mbox{\boldmath $\alpha$}}
\newcommand{\bbeta}{ \mbox{\boldmath $\beta$}}

\newcommand{\bs}{\mathbf{s}}

\newcommand{\bu}{\mathbf{u}}

\newcommand{\bw}{\mathbf{w}}

\newcommand{\bmu}{\mathbf{\mu}}

\newcommand{\bxi}{\mbox{\boldmath $\xi$}}

\newcommand{\Matern}{Mat\'{e}rn }

\documentclass[12pt]{article}
\usepackage[sort,longnamesfirst]{natbib}
\usepackage{amsbsy,amsmath,amsthm,amssymb,graphicx}
\usepackage{multirow}
\usepackage[usenames,dvipsnames]{color}
\usepackage{subfigure}
\usepackage{geometry}
\usepackage{graphicx}
\usepackage{JASA_manu}
\usepackage{longtable}

\graphicspath{{./images/}}

\begin{document}

\title{A novel principal component analysis for spatially-misaligned multivariate air pollution data}

\author{Roman A. Jandarov
        \\Division of Biostatistics and Bioinformatics
        \\ University of Cincinnati
        \\ {\tt jandarrn@ucmail.uc.edu}
            \and
        Lianne A. Sheppard
        \\Department of Biostatistics
        \\ University of Washington
        \\ {\tt sheppard@u.washington.edu}
            \and
        Paul D. Sampson
        \\Department of Statistics
        \\ University of Washington
        \\ {\tt pds@stat.washington.edu}
            \and
        Adam A. Szpiro
        \\Department of Biostatistics
        \\ University of Washington
        \\ {\tt aszpiro@u.washington.edu}
}

\date{Draft: \today}

\maketitle

\begin{center}
\textbf{Abstract}
\end{center}

We propose novel methods for predictive (sparse) PCA with spatially misaligned data. These methods identify principal component loading vectors that explain as much variability in the observed data as possible, while also ensuring the corresponding principal component scores can be predicted accurately by means of spatial statistics at locations where air pollution measurements are not available. This will make it possible to identify important mixtures of air pollutants and to quantify their health effects in cohort studies, where currently available methods cannot be used. We demonstrate the utility of predictive (sparse) PCA in simulated data and apply the approach to annual averages of particulate matter speciation data from national Environmental Protection Agency (EPA) regulatory monitors.

\vspace*{.3in}


\section{Introduction}\label{sec:intro}

One of the first well-documented air pollution events was in
Belgium in 1930. In the Meuse Valley incident a thick fog lasted
for five days which led to hundreds of people suffering from
respiratory symptoms and 60 deaths during the following days
\citep{nemery2001meuse}. Another major event was the famous London
Fog in 1952 which led to an estimated excess death toll of over
4000 \citep{logan1953mortality}. More recently, the number of
extra deaths after this event was re-estimated to be nearly
12,000. \citep{bell2001reassessment}.

Since these events, extensive research has been conducted on the
health effects of ambient air pollution exposure. A growing body
of literature presents evidence of the adverse effects of
long-term air pollution exposure on various health endpoints
\citep{samet2000fine, pope2002lung, brook2007further}. However,
many of these studies focus on a single pollutant. For example,
during recent years, much attention has been devoted to study the
role of particulate matter in air pollution, specifically
particulate matter less than 2.5 mm in aerodynamic diameter
(PM$_{2.5}$) and its effects on health \citep{miller2007long,
pope2006health}.

It is known that ambient air pollution is a complex mixture that
contains multiple pollutants. The health studies that focus on
understanding effects of a single pollutant ignore the fact that
typically people are simultaneously exposed to two or more
pollutants, and these can combine to either amplify or alleviate
their overall effects on the health endpoints. The associations
between the health endpoints and the single proxy pollutants found
in these studies are, therefore, more likely to be the effect of
the mixture (that may change by space and time), rather than the
effect of the individual pollutant
\citep{crouse2010postmenopausal}. For example, health effects of
PM$_{2.5}$ could change depending on characteristics of the
particles, including shape, solubility, pH, or chemical
composition of the pollutant \citep{vedal2012pm}. Hence, in
contrast to studies based on a single pollutant approach,
investigating health effects of multi-pollutant exposure can be
useful in epidemiological studies for two main reasons: (i) it may
provide a better understanding of potential combined effects of
individual pollutants and interactions between the pollutants and
(ii) it may result in easier interpretation of these effects to
help inform air quality management policy.

Analysis of health effects of long-term average, spatially varying
multi-pollutant air pollution exposure using cohort data has two
main challenges: (i) dimensionality of the multi-pollutant data
and (ii) the spatial misalignment of the monitoring data and
cohort study participants. The first challenge is fairly common in
any analysis of high dimensional data when there are many
independent variables of interest. In the context of understanding
health effects of exposure to multiple pollutants, dimensionality
can present difficulties because of the fact these pollutants
could potentially be highly correlated. Subsequently, this means
that estimating and interpreting parameters of interest in health
models to reveal the association between the health endpoint and
the predictors could be non-trivial.

The second challenge arises because multi-pollutant exposures are
not always available at all subject locations in a cohort study.
This challenge is obviously not unique to multi-pollutant exposure
studies. For example, in epidemiological studies of health effects
of a single pollutant, the common framework to deal with this
issue is to use prediction models to assign exposures to study
participants based on monitoring data available at different
locations. To investigate the effects of multi-pollutant exposure,
this framework requires building multiple prediction models and
assigning exposures to each pollutant at all study participant
locations. This means that if one of these prediction models
produces inaccurate predictions, the results of the health
analysis may become unreliable.

While there are methods to resolve the issues with dimensionality
and spatial misalignment in various contexts, these methods do not
deal with these challenges simultaneously. In order to understand
health effects of multi-pollutant exposures and health endpoints,
these challenges need to be solved together. In this paper, we
propose an approach for dimension reduction that can be applied to
multi-pollutant spatial data that resolves these issues in a
computationally fast unified approach. Our approach seeks to find
sparse principal component scores that explain a large proportion
of the variance in the data while also ensuring that mixtures
derived from these components are predictable at health endpoint
locations. Predictions of the lower dimensional component scores
can then be effectively used in a health effect analysis. We show
that the new approach is preferable to the sequential two-step
approach of dimension reduction followed by spatial prediction,
which may result in principal component scores that are difficult
to predict at unmeasured locations. We apply our approach to national multi-pollutant
air pollution data from the U.S. Environmental Protection Agency.

Two widely used models for predicting exposures at subject
locations are based on land-use regression (LUR) with Geographic
Information System (GIS) covariates \citep{brauer2003estimating,
hoek2008review} and universal kriging (UK)
\citep{jerrett2005spatial, kunzli2005ambient}. In addition to
using GIS covariates to construct the mean structure of the model
as in land-use regression, UK also allows for spatial dependence
by modeling correlations between residuals at different locations
\citep{kim2009health, mercer2011comparing,
sampson2013regionalized}. Since this usually results in more
accurate predictions of pollutants, we use UK throughout the
paper.

The rest of the paper is organized as follows. Section
\ref{sec:data} describes the available monitoring multi-pollutant
data and geographic covariates. Section \ref{sec:method} describes
in detail the traditional and predictive principal component
analysis approaches. Here, in Section \ref{subsec:krig}, we also
briefly give an overview of universal kriging that is used in
predicting pollutants and PC scores at new locations. In Section
\ref{sec:simulation}, we present results from our simulation
analysis to demonstrate key differences between different methods.
In Section \ref{sec:results}, we describe the application of PCA
approaches to the monitoring data. Finally, in Section
\ref{sec:discuss}, we summarize our methods and results and
discuss our statistical approach and conclusions.

\section{Data}\label{sec:data}

\subsection{Monitoring data}\label{subsec:mondata}

Air pollution data are collected nationally by a network of
monitors called the Air Quality System (AQS). Most of these data
are accessible at the Environmental Protection Agency's (EPA)
website at
http://www.epa.gov/ttn/airs/airsaqs/detaildata/downloadaqsdata.htm.
While measurements of PM$_{2.5}$ are available at almost all AQS
monitor locations, detailed data on other pollutants and
PM$_{2.5}$ components is only collected at two sub-networks of the
AQS network. One of the sub-networks of the AQS with
multi-pollutant data is called the Chemical Speciation Network
(CSN) \citep{epa2009integrated}. CSN monitoring locations are
generally selected to assess population multi-pollutant exposures,
and therefore are mostly located in urban areas. Another
sub-network of the AQS with multi-pollutant monitoring data is
called the Interagency Monitoring of Protected Visual Environments
(IMPROVE) network \citep{eldred1988improve}. IMPROVE monitoring
locations are located in national parks and some rural areas. Data
collected by the IMPROVE network are available at the Visibility
Information Exchange Web System (VIEWS), at
http://views.cira.colostate.edu/web/. IMPROVE monitoring locations
are selected to assess pollution impacts in sparsely populated
environments. Therefore, most of these monitors do not represent
exposure levels for (the majority of) subjects living in populous
areas.

The AQS monitors are located throughout the United States.
However, as funding availability and monitoring aims evolve, some
new monitors become available or/and some old ones are
discontinued, and monitoring schedules are changed. For example,
the monitors at a given location that collected PM$_{2.5}$ data
from 2000 to 2002 may not have collected PM$_{2.5}$ data from 2003
to 2005. Furthermore, an individual monitor that collected
PM$_{2.5}$ data every day in 2002 may have collected data every
third day in 2003. Another monitor that collects PM$_{2.5}$ data
every day during the winter may only collect data every third day
in the summer. Since these features complicate the analysis of the
data, we apply the methods described in this paper to annual
averages data for 2010. In 2010, the data from the AQS monitors
were at a daily resolution. These data are converted to annual
averages using completeness rules making sure that each monitor
had at least 10 data points per quarter and a maximum of 45 days
between measurements. Subsequently, before the analysis, the
annual averages were square-root transformed.

Initially, we had 7375 monitors with incomplete data for 28
pollutants: PM$_{2.5}$, PM$_{10}$, NO$_2$, NO$_x$, SO$_2$, O$_3$,
EC, OC, SO$_4$, NO$_3$, Al, As, Br, Cd, Ca, Cr, Cu, Co, Fe, K, Mn,
Na, S, Si, Se, Ni, V and Zn. Since we needed complete
multi-pollutant data for our analysis, our first goal was to
obtain a list of pollutants with a large number of monitors
without missing measurements for all pollutants in the list. After
cleaning the original data, our final list included 19 pollutants:
PM$_{2.5}$, EC, OC, Al, As, Br, Ca, Cr, Cu, Fe, K, Mn, Na, Ni, S,
Si, V and Zn. For these pollutants, the number of monitors with
full data was equal to 284. The pollutants removed from the
analysis were PM$_{10}$, NO$_2$, NO$_x$, SO$_2$, O$_3$, SO$_4$,
NO$_3$, Cd, and Co. We note that adding any of the pollutants
PM$_{10}$, NO$_2$, NO$_x$, SO$_2$ or O$_3$ to our final list
reduced the number of monitors from 284 to a number between 44 to
55, while adding any of SO$_4$, NO$_3$, Cd, or Co reduced the
number of monitors to 123. We also note that S is a good surrogate
for SO$_4$, and Cd and Co are not particularly good tracers of
major sources believed to be scientifically important, while
NO$_3$ has sampling artifact issues associated with its
volatility.

The measurements for the pollutants from the final list were from
CSN (130 monitors) and IMPROVE (154 monitors) networks. The data
from these networks for these pollutants were reported in units of
$mg/m^3$. Additionally, we note that in 2010 the CSN and IMPROVE
networks collected data using similar methods. For example, for
PM$_{2.5}$, both networks used Federal Reference Methods (FRM) to
collect data. IMPROVE data and AQS FRM data collected
concurrently at the same locations were comparable enough to be
considered equivalent. Table \ref{tablesummary} shows the list of
retained pollutants with related summary statistics. Here, based
on $R^2$s (defined as $R^2 = \mbox{max}(0,1 -
\frac{\sum(x_i-\hat{x}_i)^2}{\sum(x_i-\bar{x}_i)^2})$, where $x_i$
is the observed, $\hat{x}_i$ is the predicted values of the
pollutants) and mean squared errors (MSEs, defined as MSE =
$\frac{\sum(x_i-\hat{x}_i)^2}{n}$) obtained using cross-validation
and universal kriging (reviewed in detail in Section
(\ref{subsec:krig})), we see that while some pollutants are highly
predictable (e.g. PM$_{2.5}$ or S have high $R^2$), some
pollutants cannot be predicted well (e.g. Mn). We note here that
predictions used in calculations of $R^2$s and MSEs are obtained
from single-pollutant universal kriging models to show
descriptively how individually predictive the pollutants are.

\begin{table}
\caption{\label{tablesummary} Summary of the data for the retained pollutants from 284
monitors. All the pollutants are in units of
$mg/m^3$}\centering\tabcolsep=0.90cm
\begin{tabular*}{2.0\textwidth}{l l l l l }
\cline{1-5}\\
Pollutants & Mean & SD  & $R^2$  & MSE   \\
\cline{1-5}\\
PM$_{2.5}$ & 7.351 & 3.814 & 0.933 & 0.00156 \\
EC & 0.417 & 0.336 & 0.866 & 0.00299 \\
OC & 1.533 & 0.875 & 0.877 & 0.00729 \\
Al & 0.050 & 0.036 & 0.582 & 0.01484 \\
As & 0.001 & 0.001 & 0.817 & 0.00057 \\
Br & 0.002 & 0.002 & 0.730 & 0.00450 \\
Ca & 0.052 & 0.046 & 0.513 & 0.00001 \\
Cr & 0.001 & 0.003 & 0.497 & 0.00013 \\
Cu & 0.003 & 0.004 & 0.555 & 0.00689 \\
Fe & 0.065 & 0.068 & 0.545 & 0.00039 \\
K  & 0.051 & 0.027 & 0.668 & 0.00009 \\
Mn & 0.002 & 0.006 & 0.269 & 0.00243 \\
Na & 0.091 & 0.114 & 0.531 & 0.01473 \\
S  & 0.566 & 0.352 & 0.965 & 0.00004 \\
Si & 0.135 & 0.092 & 0.577 & 0.00477 \\
Se & 0.001 & 0.002 & 0.525 & 0.00086 \\
Ni & 0.001 & 0.001 & 0.649 & 0.00283 \\
V  & 0.001 & 0.001 & 0.783 & 0.00282 \\
Zn & 0.008 & 0.011 & 0.717 & 0.00010 \\
\cline{1-5}
\end{tabular*}
\end{table}

\subsection{Geographic covariates}\label{subsec:GIS}

The description below closely follows \cite{bergen2013national}.
For all monitor and subject locations, we obtained approximately
600 GIS covariates. These data were from various external sources,
such as TeleAtlas, the US Census Bureau, and US geological survey
(USGS). The list of GIS covariates includes distances to A1, A2,
and A3 roads (Census Feature Class Codes (CFCC)); land use within
a given buffer size; population density within a given buffer; and
normalized difference vegetation index (NDVI) which measures the
level of vegetation in a monitor's vicinity. Here, CFCC A1 roads
are limited access highways; A2 and A3 roads are other major roads
such as county and state highways without limited access
\citep{mercer2011comparing}. For NDVI, first, a series of 23
16-day composite satellite images from the year 2006 were
obtained. Then, the index was converted by the University of
Maryland from the -1 to 1 scale to the 0-255 pixel brightness
scale. On this scale, water has a value of approximately 50 and
areas with dense vegetation have values around 200. For each
location of interest, for each image, all pixels with a centroid
within a certain distance of the location were averaged (radii
included 250m, 500m, 1km, and 5km). For each buffer size, five
summary numbers were calculated from the series of 23 averages for
each location: the 25th, median, and 75th percentile of the entire
year's series, the median of the expected `high vegetation'
season, defined as April 1 - September 30, and the median of the
expected `low vegetation' season, defined as the rest of the year.

Before further analysis, all GIS covariates are pre-processed to
remove uninformative and unreliable variables with very small
variability or outliers: we eliminated variables with $>$ 85\%
identical values, and those with the most extreme standardized
outliers $>$ 7. We then log-transformed and truncated all distance
variables at 10 km and computed additional ``compiled" distance
variables such as minimum distance to major roads, distance to any
port, etc. All selected covariates are then mean-centered and
scaled by their respective standard deviations.

Additionally, we note that due to a large number of GIS covariates
(some of which are possibly correlated), using all GIS variables
directly in LUR or UK models is not practical. A commonly used
solution is to build a model by first reducing the dimensionality
of the GIS covariates into a smaller set of variables. These
extracted variables are then used to construct the mean structure
in the LUR and UK models. Various dimension reduction and variable
selection algorithms have been proposed to deal with
dimensionality of GIS covariates. These algorithms include
exhaustive search, stepwise selection, and shrinkage by the lasso
\citep[cf.][]{tibshirani1996regression, mercer2011comparing}.
Since most of these variable selection methods may be
computationally expensive, an alternative dimension reduction
approach is to apply principal component analysis to the GIS data.
Similar to analogous approaches based on partial least squares
(PLS) \citep{sampson2011pragmatic, abdi2003partial}, PCA allows us
to reduce the number of original GIS covariates by using linear
transformations into a smaller set of variables that explain most
of the variability in the GIS data (see Section \ref{sec:method}
for more on PCA). The PCA scores obtained from GIS covariates can
then be used to construct the mean structure in a LUR or UK model
instead of using all individual GIS covariates. Therefore, instead
of using all GIS covariates, we only use PCA scores obtained from
GIS data. To avoid confusion, throughout the paper, we refer to
these lower dimensional variables simply as GIS covariates. We
also note that all methods in this paper can be applied with the
full set of pre-processed GIS covariates as well.

\section{Methods}\label{sec:method}

\subsection{Review of sparse PCA for exposure dimension reduction}\label{subsec:algorithm}

We begin by reviewing dimension reduction methods (specifically
principal component analysis (PCA) and sparse PCA) and their
application to cohort studies when the exposure data are not
spatially misaligned.  We introduce notation along the way.

Suppose for each subject $i=1,\ldots,n$ in a cohort study, we
observe a scalar health endpoint $y_i$ and a $p$-dimensional
exposure vector ${\bf x}_i$. In linear regression, the most
straightforward approach to assessing associations between $y_i$
and ${\bf x}_i$ is to estimate the parameter vector ${\bbeta}$ in
\[
\label{eq:fullmodel}
{\bf Y}=\beta_0 + {\it X}{\bbeta} + {\bf \epsilon},
\]
possibly also including interactions. Here, ${\bf
Y}=(y_1,\dots,y_n)^\top$ is the vector of endpoints and ${\it X}$
is the $n \times p$ exposure matrix with rows ${\bf x}_i$.
In settings when the number of monitors/subjects is limited
compared to the number of exposure covariates and/or when there is
a complex dependence between the covariates, it can be beneficial
to reduce the dimensionality of the exposures.

A dimension reduction method identifies {low-dimensional
structure} in the exposure data ${\it X}$ and given this structure
provides a mapping from a full exposure vector to the
corresponding {low-dimensional representation}. PCA is a widely
used technique that can extract a small number of important
variables from high-dimensional data
\citep{jolliffe1986principal}. In PCA, the low-dimensional
structure is a $p\times k$ matrix $\it V$ comprised of $k$
representative $p$-dimensional exposure vectors $\{{\bf
v}_i\}_{i\leq k}$, and the low-dimensional representation is a
$n\times k$ matrix $\it U$ comprised of $n$ subject-specific
$k$-dimensional exposure vectors such that $X \sim UV^{T}$ for
$k<p$ and $X = UV^{T}$ for $k = p$. Using this representation, we
can estimate ${{\tilde{\bbeta}}}$ in
\[
\label{eq:reducedmodel}
{\bf Y}=\tilde{\beta}_0 + {\it U}{\tilde{\bbeta}} + {\bf \tilde{\epsilon}}.
\]
For $k\ll p$, the regression coefficients in ${\tilde{\bbeta}}$ are
generally easier to estimate and (may be easier to) interpret than
those in ${\bbeta }$. {We clarify that in the original regression problem, the the exposure variables could be highly correlated. This means that the components of $\bbeta$ corresponding to each exposure variable are hard to interpret in a traditional sense as the amount of change in the endpoint variable $y$ per unit increase of the exposure variable when all the other exposures are kept constant. In the exposure data, some pollutants always occur and vary together making it unrealistic to assume that pollutants can be kept constant when we change the values of other pollutants. On the other hand, the coefficients ${\tilde{\bbeta}}$ are lower dimensional than $\bbeta$ and they correspond to mixtures that are usually almost uncorrelated. This implies that the components of ${\tilde{\bbeta}}$ can be interpreted as the change in the endpoint per one unit increase of the mixture with other principal mixtures fixed.}

We focus on {unsupervised} dimension reduction methods that are not informed
by the association between $\bf Y$ and $\it X$.
An alternative is {supervised} dimension reduction,
which may sometimes identify representations that are more strongly indicative of the relationship
between $\bf Y$ and $\it X$. However, we consider it promising to work with unsupervised methods, as the resulting low-dimensional representations have the potential of being important for the multiple health
endpoints %
one faces in many epidemiology studies.

If we apply PCA to $\it X$, we find orthonormal direction or
loading vectors ${\bf v}_1, \ldots, {\bf v}_p$ and corresponding
principal component (PC) scores ${\bf u}_1 = {\it X} {\bf v}_1, \ldots, {\bf u}_p = {\it X} {\bf v}_p$. Here, dimension reduction using PCA is then
accomplished by using the first $k < p$ PC scores to represent the
data, with $k$ chosen such that these few scores and their
corresponding loadings explain most of the variation in the data
$\it X$.

In traditional PCA, all entries in the loading vectors are non-zero,
which can make it difficult to interpret individual PC scores.
Several sparse PCA methods have been proposed, with the goal of modifying PCA so that the
loading vectors contain many zeros in order to improve intepretability.
We briefly review an approach
proposed by \cite{shen2008sparse}, which exploits the connection between PCA
and low rank approximation of
matrices. For $k < p$, let us define \begin{equation} {\it X}^{(k)} =
\sum\limits_{l = 1}^{k}{\bf {u}_l} {\bf {v}_l}^\top,
\end{equation} where ${\bf {u}_l}$ and ${\bf {v}_l}$ are the first $k$ PC score and loading vectors.
It can be shown that ${\it X}^{(k)}$ is the closest rank-k matrix
approximation to $\it X$ under the Frobenius norm. For two
matrixes $\it X$ and $\it \tilde{X}$, the Frobenius norm is defined
as $\|{\it X }- {\it \tilde{X}}\|_F^2: = \mbox{tr}\{({\it X} - {\it
\tilde{X}})({\it X} - {\it \tilde{X}})^\top\}$). This implies that one
can conduct PCA by solving a series of optimization problems.
Suppose, for example, that our goal is to find the best rank-1
approximation to the matrix $\it X$. Since any $n \times p$ rank-1
matrix can be written as $\widetilde{{\bf u}} \widetilde{{\bf
v}}^\top$ with $n$-dimensional unit vector $\widetilde{{\bf u}}$
and $p$-dimensional vector $\widetilde{{\bf v}}$, we can find the
best rank-1 approximation to $\it X$ by solving the optimization
problem
\begin{equation} \label{eq:problem} \mbox{min}_{\widetilde{{\bf u}}, \widetilde{{\bf v}}} \|{\it X}
- \widetilde{{\bf u}} \widetilde{{\bf v}}^\top\|
\end{equation}
with respect to $\widetilde{{\bf u}}$ and $\widetilde{{\bf v}}$,
under the constraint $\|  \tilde{\bf u} \|=1$. We then set  ${\bf
v_1} =\widetilde{{\bf v}}/\|\widetilde{{\bf v}}\|$ and ${\bf u_1}
= X{\bf v_1}$ to obtain the first PC loading and score vectors.
Notice that we constrain $\| \tilde{\bf u}\| = 1$ in the
optimization, as this turns out to be helpful in development of
sparse PCA \citep{shen2008sparse}. We find subsequent pairs (${\bf
u}_l, {\bf v}_l$), $l > 1$ by solving corresponding one rank
approximation problems with residual matrices obtained at the
previous step.

As described in \cite{shen2008sparse}, an interpretable approach
to enforce sparsity in the loadings ${\bf v}$ can be developed via
low rank approximations. To explain the algorithm, we focus on
defining and calculating the first sparse PC score and its loading
vector. We obtain subsequent PC scores and their loadings by
simply considering the residual matrices of the sequential matrix
approximations. The main idea is to add a regularization penalty
on the loadings via Equation (\ref{eq:problem}). In other words, to find the first sparse PC loadings and scores, we solve the
minimization problem, \begin{equation} \label{eq:problem2}
\mbox{min}_{\widetilde{{\bf u}}, \tilde{{\bf v}}} \|X -
\widetilde{{\bf u}} \widetilde{{\bf v}}^\top\| +
P_{\lambda}(\widetilde{{\bf v}}),
\end{equation}
with respect to $\widetilde{{\bf u}}$ and $\widetilde{{\bf v}}$,
under the constraint $\| \tilde{\bf u} \|=1$. Here,
$P_{\lambda}({\bf y}) = \sum \lambda |y_i|$ is a $L_1$ (lasso) penalty
function with a penalty parameter $\lambda\geq 0$. This problem can
easily be solved by a fast alternating algorithm; see
\cite{shen2008sparse} for details. Similar to the above, after
$\widetilde{{\bf u}}$ and $\widetilde{{\bf v}}$ are obtained, the
first PC score is defined by ${\bf u_1} = X{\bf v_1}$ (where the
loading vector is again ${\bf v_1} =\widetilde{{\bf
v}}/\|\widetilde{{\bf v}}\|$). After iterating this procedure for
the first $k$ PCs, we define our reduced dimensional exposure
vector at monitor locations as ${\it U} =[{\bf u_1}, \ldots, {\bf
u_k}]$. In the following, we refer to this approach as traditional
sparse PCA.

We note that the penalty parameter $\lambda \geq 0$ controls the
degree of sparsity in the loadings. We will discuss various
options for selecting the penalty parameter in later sections,
when we describe our new predictive sparse PCA algorithm. If
$\lambda=0$ this algorithm recovers the standard PC loadings and
scores.

\subsection{Review of universal kriging for predicting PC scores}\label{subsec:krig}

We can apply (sparse) PCA as described above to reduce the
dimensionality of spatially misaligned exposure monitoring data.
However, our ultimate goal is to predict the PC scores at subject
locations where exposure data are not available. A straightforward
way to solve this problem is to apply (sparse) PCA without any
modification to the monitoring data and then use spatial
statistics to predict the PC scores at subject locations. We
review universal kriging (UK) in this context.

First, let ${\bf u} = \{u(\bs)\}$ be a PC score (say, the first
PC score), where $\bs \in E$, and $E$ is a set of geographical
coordinates. If $\bs$ is a monitor location, $u(\bs)$ can be
calculated. Our goal is to predict $u(\bs^\ast)$ at subject
locations $\bs^\ast$. We model ${\bf u}$ as
\begin{equation}\label{eq:linspat}
  u(\bs) = {\it Z}(\bs) \balpha + w(\bs), \mbox{ for }\bs \in E,
\end{equation}
where ${\it Z}(\bs)$ is a set of covariates associated with each
location $\bs$ and $\balpha$ is a vector of coefficients.
Traditionally, in UK, we only use the GIS covariates. Spatial
dependence is then incorporated by modeling $\{w(\bs):\bs\in E \}
$ as a zero mean stationary Gaussian process: if
$\bw=(w(\bs_1),\dots,w(\bs_n))^\top$ and $\bxi$ are the parameters
of the model, then
\begin{equation}\label{eq:lingp}
  \bw \mid \bxi \sim N(0,\it \Sigma(\bxi)),
\end{equation}
where $\it \Sigma(\bxi)$ is a symmetric and positive definite
covariance matrix of $\bw$. One commonly used covariance function is the `exponential' covariance function with parameters
$\bxi=(\psi,\kappa,\phi)$, (with $\psi,\kappa,\phi>0$), which has
the form $\it \Sigma(\bxi)_{ij} =
\psi \mathcal{I}(i=j)+\kappa \exp(-\lVert \bs_i - \bs_j \rVert
/\phi)$, where $\mathcal{I}$ is the indicator function and $\lVert
\bs_i - \bs_j \rVert$ is the Euclidean distance between locations
$\bs_i,\bs_j \in E$. This model is interpreted in the following
manner: $\psi$ (often called the nugget effect) is the non-spatial error
associated with each location, and $\kappa$ (called the sill) and
$\phi$ (called the range) are parameters that define the scale and
range of the spatial dependence respectively. {We note that throughout the paper, we use the `exponential' covariance function to model the covariance structure of the process. We also note that while there are more flexible covariance functions (e.g. \Matern function) that can be used to model the covariance matrix here, we choose to use an exponential covariance function as it know to work well with air pollution data from EPA monitors \citep[cf.][]{sampson2013regionalized, bergen2013national}. Moreover, we believe that the result and conclusions on this manuscript should be robust with regards to choosing the form of the covariance function.}

If ${\bf u}$ is the vector of observed PC scores modeled as a
Gaussian process, our goal is to obtain predictions of the score,
${{\bf u}^*}$
at new locations in $E$. Since ${\bf u}$
is a Gaussian process, the joint distribution of ${\bf u}$ and
${\bf u}^*$ given $\bxi$ and $\balpha$, the covariance and mean
parameters, respectively, can be written as:
\begin{equation}\label{eq:gausjointdist}
\begin{bmatrix} {\bf u}\phantom{a} \\ {\bf u}^* \end{bmatrix} \mid \bxi, \balpha \sim N
\left(
  \begin{bmatrix} \bmu_1\\ \bmu_2 \end{bmatrix},
\begin{bmatrix} \it \Sigma_{11} & \it \Sigma_{12} \\ \it \Sigma_{21} & \it \Sigma_{22} \end{bmatrix}\right),
\end{equation}
where $\bmu_1$ and $\bmu_2$ are the means of ${\bf u}$ and ${\bf
u}^*$ respectively and
$\it \Sigma_{11}, \it \Sigma_{12}, \it\Sigma_{21}, \it \Sigma_{22}$ are block
partitions of the covariance matrix $\it \Sigma(\bxi)$, given $\bxi,
\balpha$. $\bmu_1$ and $\bmu_2$ are functions of $\balpha$ and
$\it \Sigma_{11},\it \Sigma_{12},\it \Sigma_{21}$, and $\it \Sigma_{22}$ are
functions of $\bxi$. Consequently, using multivariate normal
theory \citep[][]{and:ims:2003}, ${\bf u}^*\mid {\bf u},
\balpha,\bxi$
is normally distributed with the mean and covariance
\begin{eqnarray}
  E({\bf u}^*|{\bf u}, \balpha,\bxi)&=&\bmu_2 + \it \Sigma_{21} \it \Sigma_{11}^{-1} ({\bf u}-\bmu_1) \label{eq:condtlref}\\ \mbox{Var}(\bu^*|\bu, \balpha,\bxi)&=&\it \Sigma_{22} - \it \Sigma_{21} \it \Sigma_{11}^{-1} \it \Sigma_{12}.\nonumber
\end{eqnarray}
To obtain predictions at new locations, we first obtain estimates
of $(\bxi,\balpha)$, $(\hat{\bxi},\hat{\balpha})$, via
maximization of the multivariate normal likelihood function
$\pi({\bf u}|\bxi,\balpha)$. Then, we use the formula for $E({\bf
u}^*|{\bf u}, \balpha,\bxi)$ in Equation (\ref{eq:condtlref}) as
predictions of ${\bf u}^*$.

\subsection{Predictive sparse PCA}\label{subsec:algorithm2}

The two stage procedure outlined in
Sections~\ref{subsec:algorithm} and~\ref{subsec:krig} can be
described as (sparse) PCA followed by spatial prediction. The main
drawback to this procedure is that there is no mechanism
in the first stage to ensure that we find scores that can be accurately predicted by a spatial model in the second stage.
Suppose there is a column of $\it X$ that accounts for a large portion of the exposure variability but
does not have much spatial structure. With complete data, we would weight this column highly in $\it V$, but with misaligned data this would result
in one or more columns of the score matrix $\it U$ being poorly predicted at subject locations. {This happens because the traditional PCA does not know that our multi-pollutant data is spatial data. Therefore, while the traditional (sparse) PCA does give more weight to pollutants with higher variability, it does not distinguish variability from spatial variability. It can be observed that some pollutants could be highly variable, while also being well informed by geographical covariates or splines. These pollutants can be predicted well by geostatistical model. At the same time, other pollutants could be highly variable and spatially unpredictable. This implies traditional (sparse) PCA may not necessarily result in scores that have better or worse predictability compared to individual predictability of the pollutants.} In this section, we extend the sparse PCA approach described in \cite{shen2008sparse} to resolve this issue by adding a constraint
to Equation (\ref{eq:problem2}) to force the PCs to be predictable
from the available data. This new algorithm is termed predictive
(sparse) PCA. We develop an alternating minimization algorithm to
solve the corresponding optimization problem. {In contrast to traditional (sparse) PCA, in predictive (sparse) PCA, we inform the algorithm that the data is spatial by including the GIS variables and spline terms. Thus, in the new PCA, we are reducing (or eliminating in some cases) the contribution of the pollutants that are highly variable, but spatially unpredictable.}

Let ${\it \tilde{Z}}(\bs)$ be a set of geographic covariates and
thin-plate spline basis functions calculated at each location
$\bs$. We will use ${\it \tilde{Z}}(\bs)$ to guide our selection of PC
loadings to ensure that they are spatially predictable. As we
noted in Section~\ref{subsec:krig}, in UK, ${\it Z}(\bs)$ does not
include splines. {We include thin-plate spline basis functions in Z(s) here so that our predictive (sparse) PCA algorithm incorporates both geographic covariates and spatial smoothing in the prediction. It is known that there exists a link between interpolation using splines and kriging \citep[cf.][]{matheron1981splines, hutchinson1994splines, dubrule1984comparing}. One can think of spline interpolation here as a low-rank kriging with fixed covariance and degree of polynomial trend to approximate the spatial variability. In other words, the spline basis functions stand in for the kriging component in identifying loading vectors with predictable scores. When it comes time to optimally predict scores at subject locations we revert back to UK with only GIS covariates in the mean model.} {In our approach, we use rank 10 thin plate splines. This is comparable to using 10 knots in conventional knot-based splines interpolation \citep[cf.][]{wood2003thin}. We select 10 because it was shown to work well in the literature for air pollution data \citep[cf.][]{bergen2013national}. In general, however, the rank of thin plate splines can be chosen by hypothesis testing approaches based on generalized cross-validation, Mallow CP criteria, maximum likelihood-based approaches. The readers are referred to \cite{wood2003thin} for more details on various methods to choose the rank of the thin-plate splines.}

Let $\it \tilde{Z}$ be the $n\times m$-dimensional matrix of the available
covariates and spline terms calculated at each monitor location.
In order to find the first principal components of the
multi-pollutant data, we propose to optimize
\begin{equation}
\label{eq:newprob} \|{\it X} - ({\it \tilde{Z}\tilde{\balpha}} / {\it ||\tilde{Z}\tilde{\balpha}||}) \widetilde{{\bf v}}^\top \|^2_F + P_{\lambda}(\widetilde{{\bf v}})
\end{equation} with respect to $\tilde{\balpha}$ and $\widetilde{{\bf v}}$. Here, we use the $L_1$ penalty function $P_{\lambda}()$ again. Note
that this optimization problem is similar to the one in Equation
(\ref{eq:problem2}). The difference is that $\widetilde{{\bf u}}$
is now constrained to be equal to ${\it \tilde{Z}\tilde{\balpha}} / {\it ||\tilde{Z}\tilde{\balpha}||}$ to force better predictability of the principal
component scores. The reason to normalize $\it \tilde{Z} \tilde{\balpha}$ is
two-fold. First, this increases identifiability of the parameters
$\tilde{\balpha}$ and $\widetilde{{\bf v}}$. Second, it allows us to
construct a fast algorithm to find a solution (see subsections
below). After optimal $\tilde{\balpha}$ and $\widetilde{{\bf v}}$ for the
first PC are obtained, since $({\it \tilde{Z}\tilde{\balpha}} / {\it ||\tilde{Z}\tilde{\balpha}||}) \widetilde{{\bf v}}^\top$ explains some part of $\bf
X$, we define the residual matrix by
\begin{equation}
\label{eq:resid}
\widetilde{\it X} = {\it X} - ({\it \tilde{Z}\tilde{\balpha}} / {\it ||\tilde{Z}\tilde{\balpha}||})\widetilde{{\bf v}}^\top .
\end{equation}
This matrix is then used to find the second PC score and its
loadings. The subsequent pairs of component scores and loading
vectors can be found by considering analogously defined residual
matrices using the $\tilde{\balpha}$ and $\widetilde{{\bf v}}$ from the
previous step.

To solve the optimization problems in Equation (\ref{eq:newprob})
with respect to $\tilde{\balpha}$ and $\widetilde{{\bf v}}$, following
ideas in \cite{shen2008sparse}, we use an iterative algorithm
described below. We first consider the following lemmas.

{\bf Lemma 1.} {For a fixed $\widetilde{{\bf v}}$, the $\tilde{\balpha}$
that minimizes Equation (\ref{eq:newprob}) has the form
$$\widehat{\tilde{\balpha}} = ({\it \tilde{Z}}^\top {\it \tilde{Z}})^{-1}{\it \tilde{Z}}^\top {\bf W},$$ where ${\bf W} = {\it X} \widetilde{{\bf v}} / \widetilde{{\bf v}}^\top\widetilde{{\bf v}}$.

{\bf Proof}. When $\widetilde{{\bf v}}$ is fixed, it is easy to
see that minimizing Equation (\ref{eq:newprob}) is equivalent to
minimizing
$$\|{\bf W} - {\it \tilde{Z}\tilde{\balpha}} / {\it ||\tilde{Z}\tilde{\balpha}||}\|^2_F.$$ which is equivalent to optimizing the following $$\|{\bf W} - {\it \tilde{Z}} \tilde{\balpha}\|^2_F
\mbox{s.t. } \|{\it \tilde{Z}}\tilde{\balpha}\|^2 = 1.$$ Hence, applying the method of
Lagrange multipliers to the latter, we obtain $$\widehat{\tilde{\balpha}} =
({\it \tilde{Z}}^\top {\it \tilde{Z}})^{-1}{\it \tilde{Z}}^\top {\bf W}.$$

{\bf Lemma 2.} {For a fixed $\tilde{\balpha}$, the optimal
$\widetilde{{\bf v}}$ that minimizes Equation (\ref{eq:newprob})
is obtained at $$\widehat{\widetilde{{\bf v}}} = h_{\lambda}({\bf
X}^\top {\it \tilde{Z}\tilde{\balpha}} / {\it ||\tilde{Z}\tilde{\balpha}||}),$$ where
$h_{\lambda}(y) = \mbox{sign}(|y|-\lambda)_{+}$.}

{\bf Proof}. First, let $w = {\it \tilde{Z}\tilde{\balpha}} / {\it ||\tilde{Z}\tilde{\balpha}||}$. Then, following \cite{shen2008sparse}, Equation
(\ref{eq:newprob}) can be minimized component-wise for each $j$ by
minimizing the following

$$\sum\limits_{i} x_{ij}^2 - 2
({\it X}^\top w)_j \widetilde{{\bf v}}_j + \widetilde{{\bf v}}^2_j +
P_{\lambda}(\widetilde{{\bf v}})$$Since
$P_{\lambda}(\widetilde{{\bf v}}) = 2\lambda \sum\limits_j
|\widetilde{{\bf v}}_j|$, the proof of the lemma then follows from
simple calculations.

We now propose the following algorithm to
solve our optimization problem:

\begin{enumerate}
\item We first choose initial values for $\tilde{\balpha}$ and $\widetilde{{\bf
v}}$.

\item For a fixed $\widetilde{{\bf v}}$, we update $\tilde{\balpha}$
using Lemma 1.

\item For a fixed $\tilde{\balpha}$, we update $\widetilde{{\bf v}}$ using Lemma
2.

\item We repeat Steps 2 and 3 until convergence.
\end{enumerate}

Suppose we are interested in the first $k$ PCs. Denote the
parameters estimated by solving Equation (\ref{eq:newprob}) k
times by $\tilde{\balpha}_1, \ldots, \tilde{\balpha}_k$ and $\widetilde{{\bf
v}}_1, \ldots, \widetilde{{\bf v}}_k$. Denote the corresponding
residual matrices by $\widetilde{X}_1, \ldots, \widetilde{X}_k$,
where $\widetilde{X}_1 = X$, $ \widetilde{X}_2 = \widetilde{X}_1 -
({\it \tilde{Z}\tilde{\balpha}_1} / {\it ||\tilde{Z}\tilde{\balpha}_1||}) \widetilde{{\bf v}}_1^\top$, etc.
For $l=1,\ldots,k$ we can now define the l-th PC loadings by ${\bf
v_l} = \widetilde{{\bf v}}_l/\|\widetilde{{\bf v}}_l\|$. Then, we
define PC scores as follows:
\begin{equation}\label{eq:way3}{\bf u}_1 = {\it X}\frac{\widetilde{{\bf v}}_1}{\|\widetilde{{\bf v}}_1\|} ={\it X} {\bf v}_1, \mbox{ } {\bf u}_2
= {\it X}\frac{\widetilde{{\bf v}}_2}{\|\widetilde{{\bf v}}_2\|}
= {\it X} {\bf v}_2 \ldots.\end{equation} As in
Section~\ref{subsec:krig} we use UK to predict the PC scores at
subject locations without exposure data, but we now expect
prediction accuracy to be improved since the loadings were
selected to ensure predictability of the scores. {Here, using the GIS covariates in predictive PCA and in UK may seem to raise suspicion with overfitting. To clarify that this is not the case, we note that the GIS covariates are not used to define the principal scores in Equation (\ref{eq:way3}). We use the GIS covariates only to guide the selection of the loadings in the predictive (sparse) PCA. Additionally, to ensure that we are not overfitting the data, we validate our approach using out of sample 10-fold cross-validation.}

\subsection{Selection of penalty parameters}\label{subsec:penalty}

We describe two methods to select penalty parameters in sparse
PCA. Recall that in both traditional and predictive approaches,
the penalty parameters, $\lambda$, control the sparsity of the loading
vectors for each PC. Since PCA algorithms are sequential, it is
easy to see that different $\lambda$ can be used for different
PCs.

The second approach is to maximize prediction
accuracy of the original pollutants by minimizing the out-of-sample Frobenius distance
$\|{\it X} - \widehat{{\it U}}{{\it V}}^\top \|_{F}$ with respect to $\lambda$ for each principal component.
{Here, $\widehat{{\it U}}$ is defined as the matrix of predictions of PC scores at monitor locations obtained via 10-fold cross-validation using UK. We use the term “out-of-sample" because these predictions are derived from applying PCA (predictive or traditional) followed by UK to each of the 10 training sets first and predicting the scores at the left out test locations. Therefore, minimizing this distance is equivalent to approximating X via PC score predictions, which requires that PC scores be predictable and that the loading vectors are not highly correlated.}

\section{Simulation Studies}\label{sec:simulation}

\subsection{Simulated Exposure Analysis}\label{subsec:sim_exposure}

In this section, we apply the methods discussed in Section
\ref{sec:method} to simulated data. Our goal is to understand the
main differences between the traditional and predictive versions
of PCA and sparse PCA. In this and the following sections, we focus
on the first three principal components. {We use the proportion of variance accounted for using the principal scores and the consideration of simplicity in selecting the number of components in the analysis. With three principal scores, we are able to explain about 70-80\% of the original variance in the simulated and real datasets in Section \ref{sec:results}. The threshold values of 70-80\% are often used as a rule of thumb in the PCA literature \citep[cf.][]{miller1998handbook, kim1978factor}. In our manuscript, three is also small enough to be useful in presenting the material without creating excessively large tables. We note that using the amount of explained variance in the data with lower dimensional scores is not the only way to select the number of components in the analysis. For example, other commonly used methods include approaches based Kaiser criterion \citep{kaiser1960application} and the scree tests \citep{cattell1966scree}. These and various other statistical and non-statistical methods and statistical test to find optimal number of components are discussed in detail in \cite{peres2005many, bryant1995principal} and \cite{Gorsuch1983principal}.}

We first randomly choose 400 locations from the list of 970
monitor locations. At these locations, we have access to all GIS
covariates. Using the notation from Section
\ref{subsec:algorithm2}, let $\it \tilde{Z}$ be the matrix of the GIS
covariates and thin-plate splines obtained using the coordinates
of the selected locations. Using $\it \tilde{Z}$, we generate
19-dimensional multi-pollutant data, $\it X$, by simulating its
columns ${\it X}[,j]$, using linear models of the form
$${\it X}[,j] = {\it \tilde{Z}}\gamma_j + {\boldmath \epsilon}_j,$$ where $j = 1,..., 19$, with $\bf \gamma_j$ fixed coefficient vectors and i.i.d. noise vectors ${\boldmath \epsilon}_j$ distributed
according to a multivariate normal distribution with mean 0 and covariance
matrix $\sigma_j \it I$.

We choose the simulation parameters $\gamma_j$ and $\sigma_j$ so
that some pollutants depend only on GIS
covariates; some pollutants depend only on thin-plate splines and
some pollutants are just noise. In addition, we also generate some
pollutants using a combination of GIS covariates and splines. As
can be seen from the model above, all of this is easily achieved
by making some of the $\gamma_j$'s close to zero. Finally, we also
introduce correlation between pollutants by forcing some of the
$\gamma_j$'s to be close to each other for different pollutants in
order to make the simulated data reflect some of the correlation
structure observed in the actual monitoring data.

Additionally, we consider two different scenarios. Under the first
scenario, we choose the parameters in the data generating model so
that most of the simulated pollutants are predictable using UK.
Under the second scenario, we make sure that most of the
pollutants cannot be predicted well using a spatial model. {We note that this can be achieved by varying the values of the coefficients $\gamma_j$ and the variances of the error terms $\sigma_j$: increasing the values of $\gamma_j$ corresponding to the GIS covariates increases the predictability of the pollutants using the geographical features of the locations, which in turn increases the predictability using UK since GIS variables are a part of the kriging model; increasing the values of $\gamma_j$ corresponding thin-plate splines results in smoother pollutant surfaces, which increases predictability using UK since kriging captures spatial variability; and finally, adding random noise with larger variances $\sigma_j$ decreases predictability by making it hard to detect the true signal.}

\begin{table}
\caption{\label{table:sim1} Summary: Simulated data with pollutants with high
predictability. The $R^2$ is obtained using UK. Absolute
correlation is defined as the average of three absolute
correlations between the scores. Sparseness is the fraction of
zero-valued loadings. Here, ``Max.Scores'' means that the penalty was selected to maximize the predictability of the principal scores; and ``Max.Pollutants'' means that the penalty  was selected to maximize the predictability of the pollutants}\centering \tabcolsep=0.23cm
\begin{tabular*}{2.0\textwidth}{l l l l l l l l }\cline{1-8}\\
{   } & {   } & {   } &
\multicolumn{3}{c}{\parbox[c]{3cm}{Predictability of Scores
($R^2$)}} & \multirow{2}{*} {\parbox[t]{3.5cm}{Abs.
Correlation\\(Average)}} & \multirow{2}{*} {Sparseness (\%)}\\
\cline{4-6}\\
{   } & {   } & {   } & PC1 & PC2 & PC3\\

\cline{1-8}\\

\multicolumn{2}{c}{\multirow{2}{*}{\rotatebox[origin=rB]{90}{\parbox[c]{0.8cm}{Without
\\ Penalty}}}} & Trad. PCA &
0.96 & 0.87 & 0.70 & 0.04& 0.00\%\\
& & Pred. PCA &
0.96 & 0.88 & 0.71 & 0.05& 0.00\%\\[0.5cm]

\cline{1-8}\\

\multirow{4}{*}{\rotatebox[origin=c]{90}{\parbox[c]{3cm}{With
Penalty}}} &
\multirow{2}{*}{\rotatebox[origin=c]{90}{\parbox[c]{1cm}{Max.
Scores}}}& Trad. PCA &
0.97 & 0.91 & 0.84 & 0.38& 35.22\%\\
& & Pred. PCA &
0.97 & 0.92 & 0.84 & 0.43& 35.53\%\\[0.5cm]

\cline{2-8}\\

& \multirow{2}{*}{\rotatebox[origin=c]{90}{\parbox[c]{1.45cm}{Max.
\\ Pollutants}}}&
Trad. PCA &
0.96 & 0.87 & 0.66 & 0.11& 19.96\%\\
& & Pred. PCA &
0.96 & 0.87 & 0.74 & 0.18& 20.04\%\\[0.5cm]

\cline{1-8}
\end{tabular*}
\end{table}

\begin{table}
\caption{\label{table:sim2} Summary: Simulated data with pollutants with low
predictability. The $R^2$ is obtained using UK. Absolute
correlation is defined as the average of three absolute
correlations between the scores. Sparseness is the fraction of
zero-valued loadings. Here, ``Max.Scores'' means that the penalty was selected to maximize the predictability of the principal scores; and ``Max.Pollutants'' means that the penalty was selected to maximize the predictability of the pollutants}\centering\tabcolsep=0.23cm
\begin{tabular*}{2.0\textwidth}{l l l l l l l l }\cline{1-8}\\
{   } & {   } & {   } &
\multicolumn{3}{c}{\parbox[c]{3cm}{Predictability of Scores
($R^2$)}} & \multirow{2}{*} {\parbox[t]{3.5cm}{Abs.
Correlation\\(Average)}} & \multirow{2}{*} {Sparseness (\%)}\\
\cline{4-6}\\
{   } & {   } & {   } & PC1 & PC2 & PC3\\

\cline{1-8}\\

\multicolumn{2}{c}{\multirow{2}{*}{\rotatebox[origin=rB]{90}{\parbox[c]{0.8cm}{Without
\\ Penalty}}}} & Trad. PCA &
0.76 & 0.60 & 0.27 & 0.02& 0.00\%\\
& & Pred. PCA &
0.85 & 0.76 & 0.56 & 0.08& 0.00\%\\[0.5cm]

\cline{1-8}\\

\multirow{4}{*}{\rotatebox[origin=c]{90}{\parbox[c]{3cm}{With
Penalty}}} &
\multirow{2}{*}{\rotatebox[origin=c]{90}{\parbox[c]{1cm}{Max.
Scores}}}& Trad. PCA &
0.93 & 0.79 & 0.28 & 0.31& 52.47\%\\
& & Pred. PCA &
0.95 & 0.90 & 0.83 & 0.42& 69.96\%\\[0.5cm]

\cline{2-8}\\

& \multirow{2}{*}{\rotatebox[origin=c]{90}{\parbox[c]{1.45cm}{Max.
\\ Pollutants}}}&
Trad. PCA &
0.79 & 0.66 & 0.28 & 0.08& 19.14\%\\
& & Pred. PCA &
0.88 & 0.80 & 0.61 & 0.13& 25.69\%\\[0.5cm]

\cline{1-8}
\end{tabular*}
\end{table}

At each simulation, we have 400 locations with multi-pollutant
data. In assessing the performance of different models, we split
the simulated data into two parts with 300 and 100 locations. We
apply the methods to data from 300 locations. Then, we predict the
pollutants and scores at the remaining 100 location that were not
used in building the model. We compare these predictions to known
values of the pollutants and scores. We note here that the known
scores are defined by projecting the observed values of pollutants
at 100 locations to fixed loadings obtained from applying PCA to
data from 300 monitors.

Tables \ref{table:sim1} and \ref{table:sim2} summarize results
from the simulation study under different scenarios. In these
tables, simulated examples are run 100 times and all summaries are
based on averages. The ordering of principal component scores is
from the highest predictable to the lowest predictable score in terms of $R^2$. In
other words, at each simulation, the best predictable PC score is
called the first score, the second best predictable score is
called the second score etc. Here, we define the absolute
correlation as an average of three absolute correlations between
all three scores, and sparseness is the fraction of loadings equal
to zero. Here, since correlations with different signs can result
in an average close to zero, looking at the average of absolute
correlations is a more informative summary of dependence in
scores.

Table \ref{table:sim1} presents the first simulation scenario
where all pollutants are highly predictable using a spatial model.
We see that traditional and predictive PCA perform similarly when
there is no penalty and when the sparseness penalty parameters are
chosen to maximize predictability of the scores. However, when the
penalty parameters are chosen to maximize predictability of the
pollutants, we see that the predictive method increases the $R^2$
for the third principal score compared to traditional sparse PCA.

Table \ref{table:sim2} presents the second simulation scenario
where only some pollutants are highly predictable using a spatial
model. We expect to see more of a benefit from predictive
(sparse) PCA in this setting, and indeed the advantage is more
pronounced regardless of how the penalty is chosen. With no
penalty, predictive PCA increases predictability of the PC scores
without increasing the correlations between the scores. When the
penalty is chosen to maximize predictability of scores, the
average correlation between the scores tends to be undesirably
high. When the penalty is selected to maximize predictability of
pollutants, both the traditional and predictive sparse PCA methods
result in relatively small correlations between its scores, but
prediction accuracy is better for predictive sparse PCA.

\subsection{Simulated Health Analysis}

{While it is possible and straightforward to directly apply the
methods described in this manuscript to health data from cohort
studies to investigate the effects of multi-pollutant mixtures on health
endpoints, to obtain reliable health effects estimates, the health
analysis should also address additional challenges like accounting
for measurement error \citep{szpiro2013measurement}, effects of preferential
sampling of monitoring locations \citep{lee2015impact}, and possible effect
modifiers in the health model \citep{analitis2014effects, delfino2014asthma, park2008air}. We do not aim to present the
results of an analysis of real cohort studies here, which is a topic
for future work. Instead, in this section, we provide a simulated
health analysis example of the effects of exposure to multiple pollutants
using traditional and predictive sparse PCA.

We follow the idea laid out in Section \ref{subsec:sim_exposure} and assign
$N_1 = 300$ locations as monitors locations and $N_2 = 7075$ locations for
our cohort subjects. We select these locations from the full list
of available 7375 EPA monitoring sites for convenience so that we
do not have to recalculate the geographical covariates. We note
that in Section \ref{subsec:sim_exposure}, we used the list of EPA AQS sites
without missing data, but this does not affect the results of this
analysis. We also note that we intentionally make the number of
monitor locations small (300), compared to the number of subject
locations in our simulated cohort data (7075), as this is usually
the case in real cohort studies where sizes of cohorts are much
larger than the number of the monitoring sites. We generate
the true exposure surface measurements at $N_1+N_2$ locations using the methods
described in Section \ref{subsec:sim_exposure}. For shortness sake, however, we
only consider the second scenario where we made most of the
pollutants spatially unpredictable. From Section \ref{subsec:sim_exposure}, we
know that the predictive PCA results in principal scores that are
predictable better compared to the scores from traditional PCA.
Our primary goal here is to see how this affects the estimated health effects in the health analysis. After assuming that our simulated
subjects live at the selected $N_2$ addresses, for each subject,
we generate 5 subject specific covariates. These are A = age, R = race,
I = income, E = education, and S = smoking status. We assume that age is a
uniform continues variable between 30 and 80; and race, income,
education and smoking status are all categorical variables with 4
categories, with realistic proportions. We assume that we are
interested in 2 health endpoints, $Y_1$ and $Y_2$. For example,
the first endpoint $Y_1$ could be a cardiovascular health
endpoint; the second endpoint $Y_2$ could be a respiratory endpoint.
In the true health model, we use the following to generate the
health endpoints:}

\begin{equation}
\label{eq:true_health_model1}
  Y_1 =    0.35* \mbox{P}_1 + 0.5*
  \mbox{A} + 2*\mbox{R}
- 0.5* \mbox{I} - 1* \mbox{E} + \mbox{S} +
\mbox{Error}\end{equation}

\begin{equation}
\label{eq:true_health_model2}Y_2 =    0.6* \mbox{P}_5 +
0.6* \mbox{P}_8 + 0.5*\mbox{A} + 2*\mbox{R} - 0.5*
\mbox{I} - 1* \mbox{E} + \mbox{S} +
\mbox{Error},\end{equation} {where we assume the error terms are normally distributed with standard deviation equal to 10. Here, $\mbox{P}_1$, $\mbox{P}_5$ and $\mbox{P}_8$ are the simulated pollutants, and age, race, income, education, smoking status are denoted by A, R, I, E, and S respectively.

Finally, assuming that the true pollutant measurements are only
available at the monitor locations, we want to study the effects
of multi-pollutant exposure on $Y_1$ and $Y_2$ using the traditional and
predictive (sparse) PCA. When we apply PCA approaches to simulated
monitoring data, we only look at the first 3 principal components again.
As noted before, with 3 components, we are able to explain approximately 70\% of
the original variability in the data. A more thorough approaches
and strategies exist to select the number of components to
consider in the analysis. Selecting the number of components using
these approaches would be worthwhile but is beyond the scope of
this paper. For each component, we choose the penalty parameters
to maximize predictability of the pollutants. We see from the
previous section that this yields scores that are predictable and
not highly correlated compared to other methods to choosing the
penalty parameters in both traditional and predictive PCA. Once
predictions of the principal scores are obtained, we fit a
regression model with principal scores as main predictors. In this
model, we also include all the subject-specific covariates. In
fitting this model, we do not expect to recover the exact true
parameters and variables used to generate the data in the true
model in Equations \ref{eq:true_health_model1} and
\ref{eq:true_health_model2}. In contrast to variable selection
methods, both the predictive and traditional PCA are unsupervised
methods to extract mixtures of pollutants that occur together.
When we study the effect of these mixtures on our health endpoints
using predictions of the scores, we want to see if the the scores
are significantly associated with the endpoints and if these
mixtures contain the true ``bad actors'' used in generating these
data.

Results from one run of this analysis are given in Tables
\ref{table:sim_interp} and \ref{table:sim_health_results}. First,
from Table \ref{table:sim_interp}, we see how well we can predict
the principal scores from traditional and predictive PCA at
subject locations. Here, to calculate these $R^2$s, we use the
predicted and the true values of the scores. The results of this table
are consistent with what we see in Section \ref{subsec:sim_exposure}. In this table,
we also see the interpretations of the principal scores derived
from the sparse loadings. For example, the first principal score
is mostly a mixture of pollutants 1, 2, 4, 9, and 14.

In Table \ref{table:sim_health_results}, we present the results of the health analysis. In Table \ref{table:sim_health_results}, for $Y_1$, at the significance level of 0.05, we see that PC2 from predictive PCA is associated with $Y_1$. From Table \ref{table:sim_interp}, we see that an important part of this mixture is the pollutant 1, which we know is the true predictor of $Y_1$ in Equation \ref{eq:true_health_model1}. Here, we also see that the PC1 from traditional PCA (which also contains the pollutant 1) is marginally significant. For $Y_2$, for predictive PCA, we see that the PC3 is associated with $Y_1$. From Table
\ref{table:sim_interp}, this is a mixture that contains the pollutant 8 which was used in Equation \ref{eq:true_health_model2}. We see that the PC2 (which also contains the pollutant 5) is marginally significant. The pollutant 5 is also used in the true model in Equation \ref{eq:true_health_model2}. For $Y_2$, for traditional PCA, we see that the PC3 is associated with $Y_1$. The PC3 here is a mixture that contains the pollutant 8. From this example, it is clear that the predictive PCA outperforms the traditional PCA when the goal is to study the effects of principal mixtures on the health endpoints. In our example, we observe this for both $Y_1$, where the signal for the score from predictive PCA was stronger, and for $Y_2$, where the traditional PCA missed the did not capture the Pollutant 5. We note that when we ran the simulated health analysis multiple times, the predictive PCA preformed better consistently.}

\begin{table}
\caption{\label{table:sim_interp} Interpretation of simulated principal scores}
\centering\tabcolsep=0.10cm
\begin{tabular*}{3.0\textwidth}{l l l }
\cline{1-3}\\
\multirow{3}{*}{\parbox[c]{3cm}{Traditional sPCA}} & PC1 ($R^2 = 0.6$)& Mostly a mixture of pollutants 1, 2, 4, 9, 14 \\
           & PC2 ($R^2 = 0.6$)& Mostly a mixture of pollutants 1, 7, 11, 12, 13, 15, 17\\
           & PC3 ($R^2 = 0.6$)& Mostly a mixture of pollutants 6, 7, 8, 10, \\[0.5cm]

\multirow{3}{*}{\parbox[c]{3cm}{Predictive sPCA}} & PC1 ($R^2 = 0.6$)& Mostly a mixture of pollutants 1, 2, 4, 9, 14 \\
           & PC2 ($R^2 = 0.6$)& Mostly a mixture of pollutants 1, 2, 4, 5, 9, 14, 17 \\
           & PC3 ($R^2 = 0.6$)& Mostly a mixture of pollutants 6, 7, 8, 10\\
\cline{1-3}
\end{tabular*}
\end{table}

\begin{table}
\caption{\label{table:sim_health_results} Inference for simulated health model coefficient}
\centering \tabcolsep=0.45cm
\begin{tabular*}{3.0\textwidth}{l l l l l l}
\cline{1-6}\\
 & & \multicolumn{2}{c}{Traditional sPCA} &  \multicolumn{2}{c}{Predictive sPCA}\\
 \cline{3-4}
 \cline{5-6}
 & & beta & pval & beta & pval\\[0.5cm]

\multirow{3}{*}{\parbox[c]{3cm}{Endpoint 1}} & PC1 & -0.29 & 0.08 & -0.01 & 0.89\\
           & PC2 & -0.04 & 0.66 & 2.07 & 0.04\\
           & PC3 & -0.02 & 0.80 & 0.18 & 0.86 \\[0.5cm]

\multirow{3}{*}{\parbox[c]{3cm}{Endpoint 2}} & PC1 & -0.27 & 0.11 & -0.36 & 0.72 \\
           & PC2 & -0.06 & 0.53 & 1.83 & 0.07 \\
           & PC3 & -0.22 & 0.02 & 2.24 & 0.03\\
\cline{1-6}
\end{tabular*}
\end{table}

\section{Application to multi-pollutant monitoring data}\label{sec:results}

We now apply traditional and predictive (sparse) PCA to the
multi-pollutant monitoring data described in Section
\ref{sec:data}. As an exploratory analysis, in Section
\ref{sec:data}, Table \ref{tablesummary}, we observed that while
some pollutants are highly predictable, there are others that have
poor cross-validated $R^2$. This suggests that PC scores obtained
from projecting the multi-pollutant data onto traditional PC
loadings may not have good spatial predictability.

\begin{figure}
\caption{Scatterplots of the predictions of the scores without a
penalty. These are obtained using cross-validation and
UK}\centering
    \begin{center}
    \includegraphics[scale=1.4]{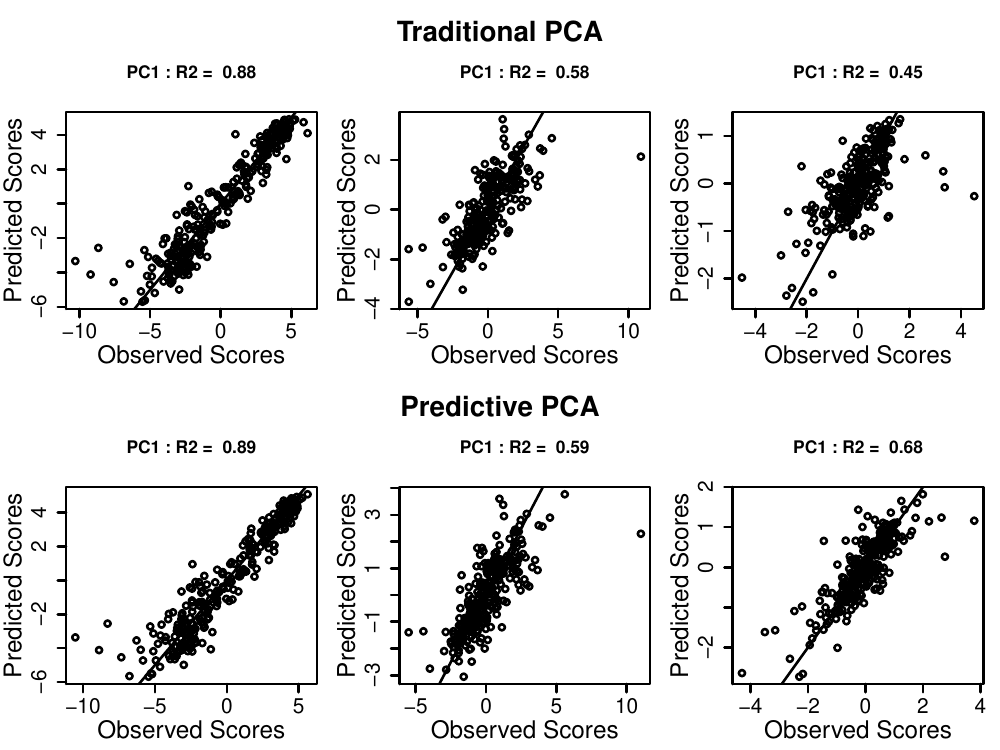}
    \end{center}
\label{figure:scatter1}
\end{figure}

\begin{figure}
\caption{The principal component loadings from two PCA approaches without penalty}\centering
    \begin{center}
    \includegraphics[scale=0.5]{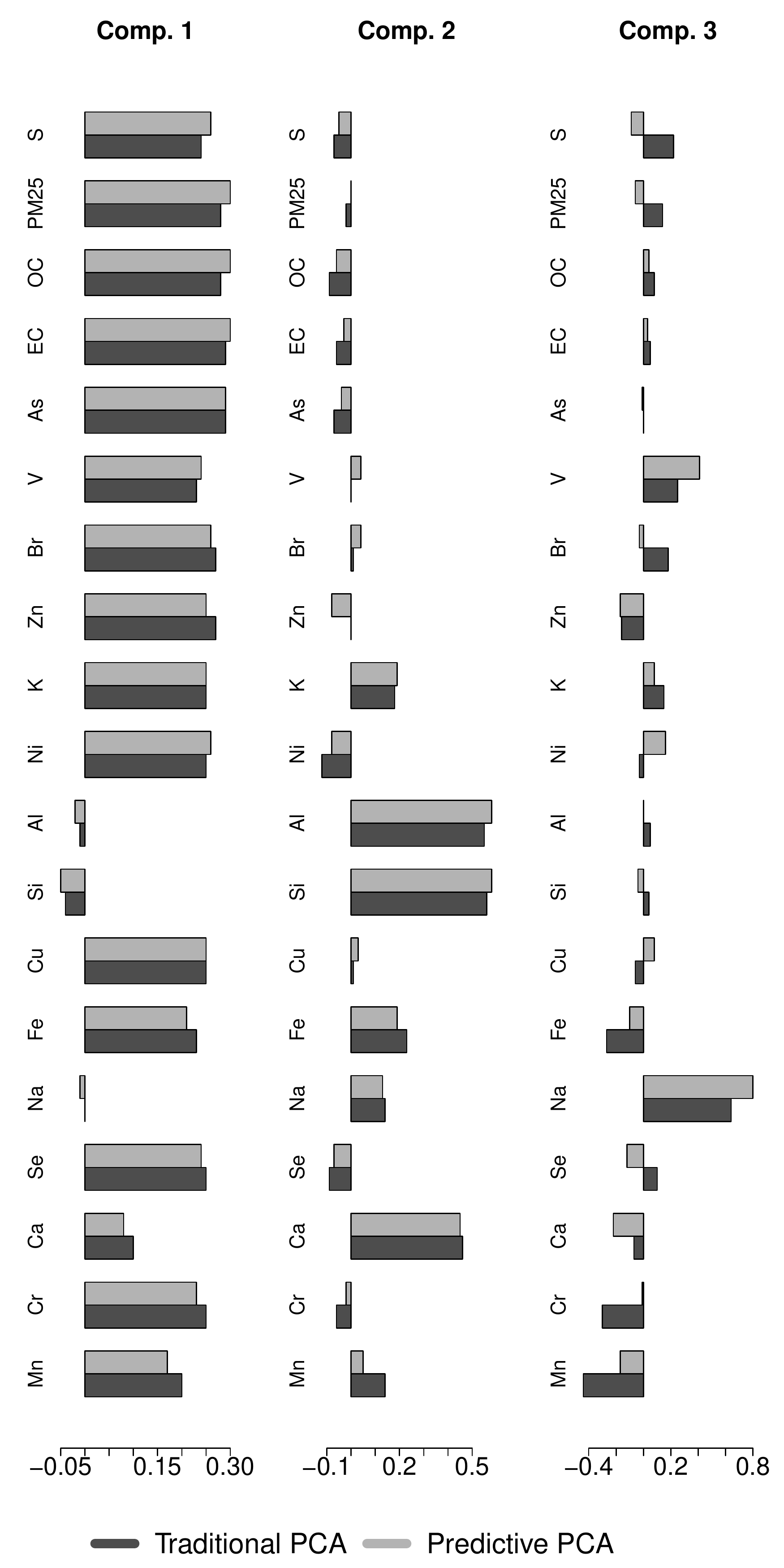}
    \end{center}
\label{figure:load1}
\end{figure}

First, we apply traditional and predictive PCA without a penalty.
Figure \ref{figure:scatter1} shows scatterplots obtained from
10-fold cross-validation. From this figure, we conclude that the
cross-validated $R^2$ for the PC1 and PC2 scores from the two
methods are very similar. This is also clear from looking at the
bar plot of the corresponding loadings in Figure
\ref{figure:load1}, where we see that loadings from traditional
and predictive PCA for the first two principal component scores
are close to each other. We note, however, that predictive PCA
increases the predictability of the PC3 scores from $0.45$ to
$0.68$. From Figure \ref{figure:load1}, where pollutants are
ordered from the smallest individual $R^2$ to the highest, we see
that the different PCA methods have different loadings for PC3.

\begin{table}
\caption{\label{table:results} Summary of results. The $R^2$ is obtained using UK.
Sparseness is the fraction of zero-valued. Here,  ``Max.Scores'' means that the penalty was selected to maximize the predictability of the principal scores; and ``Max.Pollutants'' means that the penalty was selected to maximize the predictability of the pollutants loadings}\centering
\tabcolsep=0.11cm

\begin{tabular*}{2.0\textwidth}{l l l l l l l l l l
} \cline{1-10}\\[0.1cm]

{   } & {   } & {   } & \multicolumn{3}{c}
{\parbox[c]{3cm}{Predictability of Scores($R^2$)}} &
\multicolumn{3}{c} {\parbox[c]{3.5cm}{Abs.
Correlation\\(Average)}} & \multirow{2}{*} {Sparseness (\%)}\\[0.1cm]

\cline{4-6} \cline{5-9}\\[0.1cm]

{   } & {   } & {   } & PC1 & PC2 & PC3 & PC1vsPC2 & PC1vsPC3 & PC2vsPC3\\[0.1cm]

\cline{1-10}\\

\multicolumn{2}{c}{\multirow{2}{*}{\rotatebox[origin=c]{90}{\parbox[c]{1.3cm}{Without
\\ Penalty}}}} & Trad. PCA &
0.88 & 0.58 & 0.45 & 0.00& 0.00& 0.00& 0.00\%\\
& & Pred. PCA &
0.89 & 0.59 & 0.67 & -0.05& 0.03& -0.04& 0.00\%\\[0.5cm]

\cline{1-10}\\

\multirow{4}{*}{\rotatebox[origin=c]{90}{\parbox[c]{3cm}{With
Penalty}}} &
\multirow{2}{*}{\rotatebox[origin=c]{90}{\parbox[c]{1cm}{Max.
Scores}}}& Trad. PCA &
0.92 & 0.88 & 0.64 & 0.96& -0.14& -0.08& 54.39\%\\
& & Pred. PCA &
0.93 & 0.90 & 0.89 & 0.97& 0.97& 0.99& 43.86\%\\[0.5cm]

\cline{2-10}\\

& \multirow{2}{*}{\rotatebox[origin=c]{90}{\parbox[c]{1.45cm}{Max.
\\ Pollutants}}}&
Trad. PCA &
0.88 & 0.59 & 0.57 & 0.17& -0.15& -0.19& 47.37\%\\
& & Pred. PCA &
0.89 & 0.59 & 0.66 & -0.04& 0.01& 0.01& 17.54\%\\[0.5cm]

\cline{1-10}
\end{tabular*}
\end{table}

Now we apply both traditional and predictive PCA
with a penalty to induce sparseness. A summary of the results from this analysis is given in Table
\ref{table:results}. {If we compare this table to Tables \ref{table:sim1} and \ref{table:sim2}, we notice that Table \ref{table:results} contains information about the exact correlations between the principal scores, whereas in Tables \ref{table:sim1} and \ref{table:sim2}, we only present the average absolute correlations between the scores. We do this because in simulated examples, the definition of the order of the principal components is different. In other words, the role of PC1, PC2 and PC3 changes from simulation to simulation making it confusing to interpret the meanings of the of the individual correlations.} In Table \ref{table:results}, sparseness is again defined as the fraction of zeros in the PC loadings vector. Sparseness is equal to zero for
non-sparse PCA for both traditional and predictive PCA, as can be
seen in the first part of Table \ref{table:results}. When the penalty is chosen to maximize predictability of the scores, we see
that the cross-validated $R^2$'s for all PC scores are high for
both methods. In sparse predictive PCA, these $R^2$'s are above
$0.90$. However, we note that both methods result in scores that
are highly correlated.

{When the penalty parameter is chosen to maximize predictability of
the pollutants, Figure \ref{figure:trad1} and Table \ref{table:results} show that while the PC1
and PC2 scores from the traditional and predictive versions of
sparse PCA have similar $R^2$'s, the predictability of the PC3
score from the predictive method is higher than from the
traditional method.  The bar plot of the corresponding principal loadings for the predictive and traditional sparse PCA are given in Figure \ref{figure:load2}. Here, we see that the noticeable differences in
the loadings are again in PC3. As expected, we also notice that the
loadings now are sparse.} From these, it is also clear that the predictive sparse PCA
results in lower correlations between scores when compared to the
traditional method, which is consistent with the conclusions
obtained in simulated examples in Section \ref{sec:simulation}. {Additionally, we see that the traditional PCA has higher sparseness compared to the predictive PCA in Table \ref{table:results}. However, we do not believe that is something that is systematic between the methods as we did not observe the phenomena that sparseness was substantially different for two methods in simulated examples.  Looking at Figure \ref{figure:load2}, we see that the difference in sparseness of the loadings in Table \ref{table:results} is mostly because of the differences in the third principal loadings. We speculate that while the loadings values for positive weights are similar, the predictive PCA is adding elements to the mixture with negative loadings values to increase the overall predictability of the final principal score. This elements are given a loading values equal to zero in the traditional PCA.} {In Figure \ref{figure:heat}, we present a heatmap of the predictions of the scores from predictive sparse PCA. These heatmaps show clearly how different mixtures derived from different scores vary spatially} {When we apply universal kriging model to predict the sparse principal scores in Table \ref{table:results}, the estimated range for all the scares was around 600km for both traditional and predictive sparse PCA. The estimate of the partial sill and nugget effect for the first principal score was 40 and 0.92 for traditional PCA, and 37 and 0.82 for predictive PCA; for the second principal score, 79 and 0.76 for traditional PCA, and 73 and 0.81 for predictive PCA; and for the third principal score, 69 and 0.21 for traditional PCA, and 65 and 0.19 for predictive PCA respectively.}

\begin{figure}
\caption{The principal component loadings from two PCA approaches with penalty}\centering
    \begin{center}
    \includegraphics[scale=0.5]{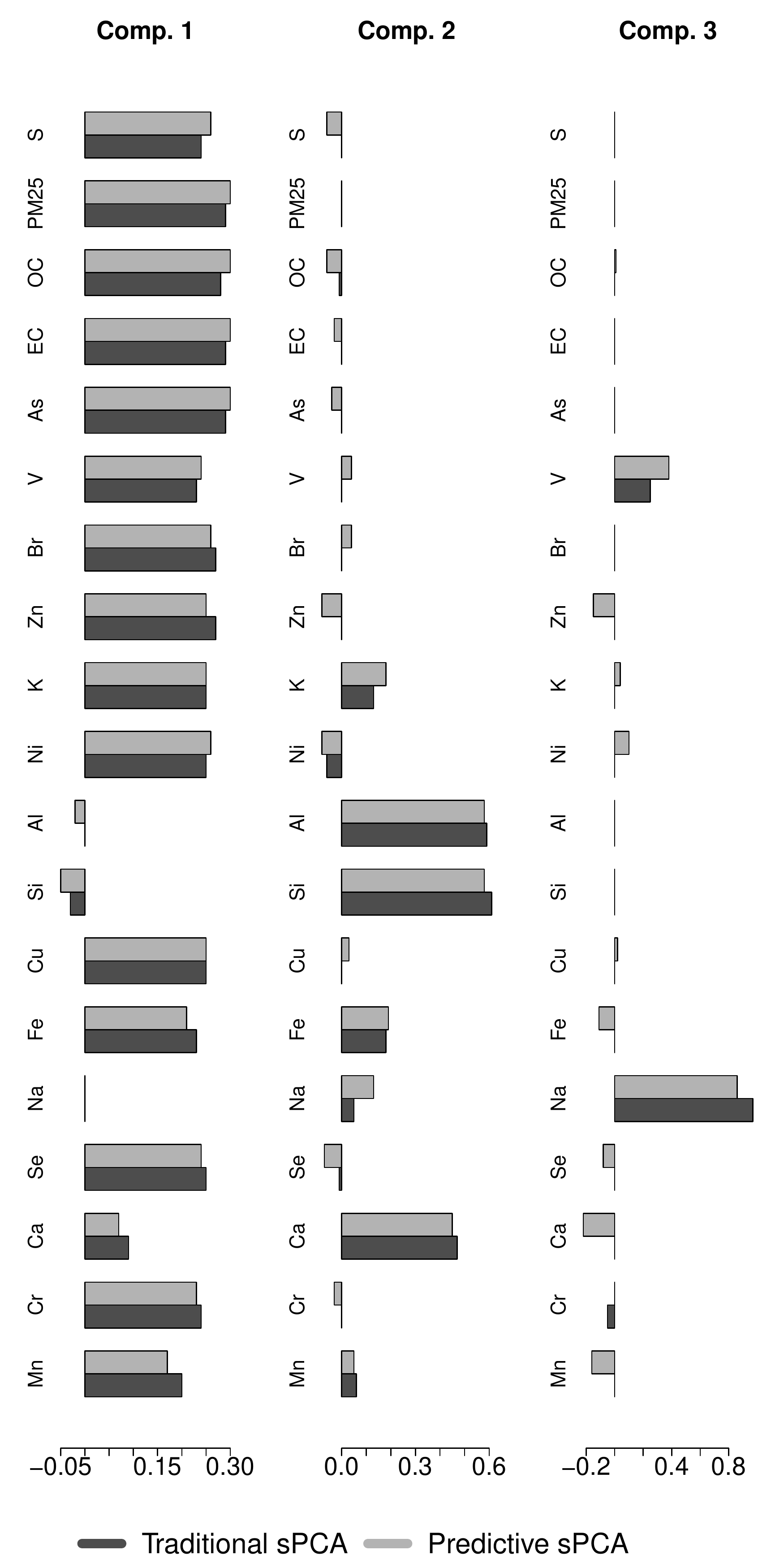}
    \end{center}
\label{figure:load2}
\end{figure}

\begin{figure}
\caption{Scatterplots of the predictions of the scores with a
penalty to maximize predictability of the pollutants}\centering
    \begin{center}
    \includegraphics[scale=1.4]{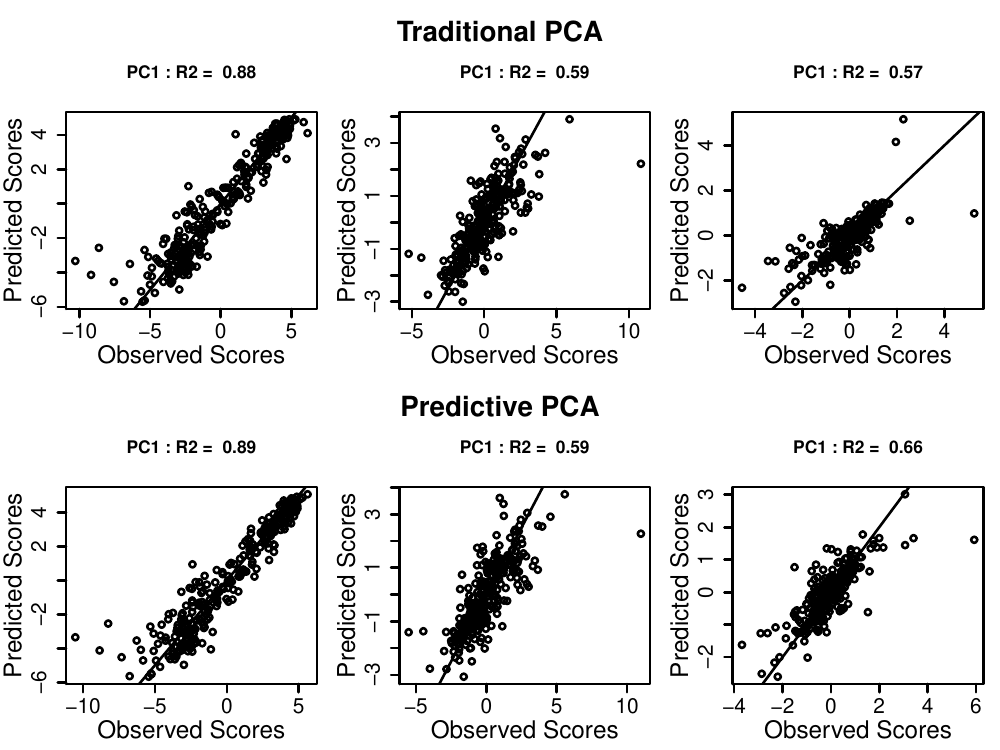}
    \end{center}
\label{figure:trad1}
\end{figure}

\begin{figure}
\caption{Heatmaps based on the predictions of the scores from
predictive sparse PCA }\centering
    \begin{center}
    \includegraphics[scale=1]{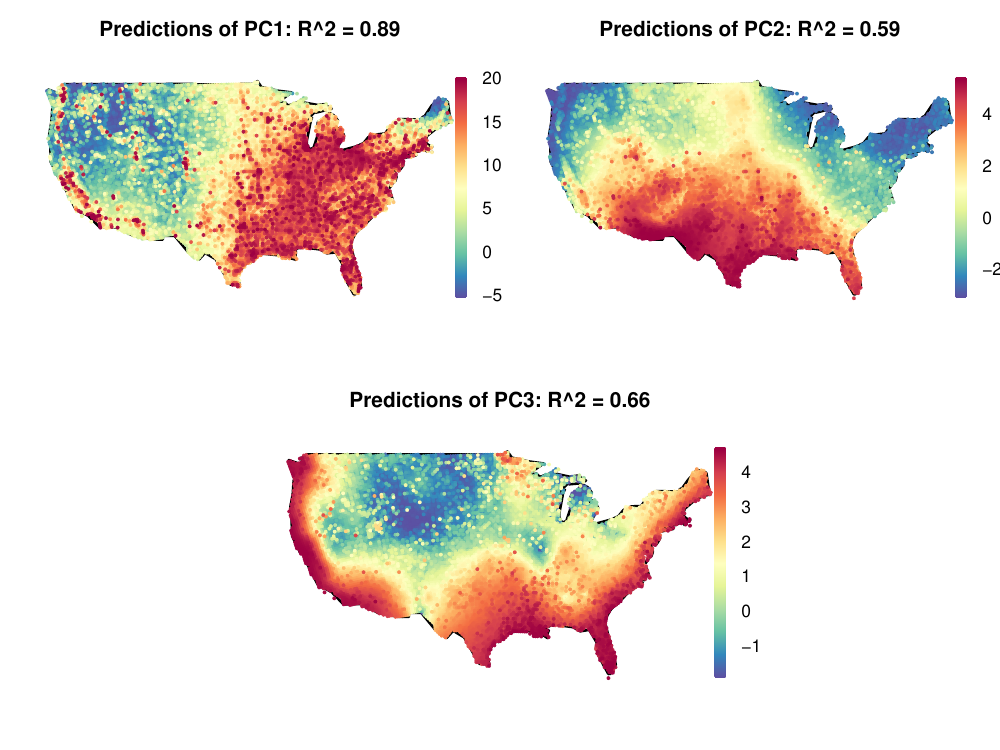}
    \end{center}
\label{figure:heat}
\end{figure}

\section{Discussion}\label{sec:discuss}

In this paper, we describe a novel predictive (sparse) PCA
approach that can be used to obtain exposure estimates in air
pollution cohort studies. The approach can be used with or without
a penalty. In contrast to traditional PCA, we demonstrate that
predictive PCA may lead to principal scores that can be predicted
well using geographical covariates and splines in the exposure
prediction stage. We have also shown that penalty parameters in
predictive sparse PCA can be selected so that the correlations
between the resulting scores are small. Based on the simulated
examples and data analysis, we observed that when the penalty
parameter is selected to maximize predictability of the principal
scores, the resulting loadings are nearly orthogonal and the
scores are almost uncorrelated. {In our data analysis, we apply the traditional and predictive (sparse) PCA approach to multi-pollutant data from EPA's regulatory monitors from 2010. In Table \ref{tablesummary}, we present a summary of how individual pollutants can be predicted using the commonly used universal kriging model with GIS covariates. In Table \ref{table:results} and Figures \ref{figure:scatter1} and \ref{figure:trad1}, we show how the different mixtures obtained from the principal scores can be predicted. In Figures \ref{figure:load1} and \ref{figure:load2}, we present the component loadings of the principal scores obtained using the predictive and traditional (sparse) PCA. Finally, Figure \ref{figure:heat} provides further insights on how the principal component scores vary spatially across the U.S.}

In predictive (sparse) PCA, we do not use the health data when
applying the algorithm to multi-pollutant data. Therefore, one may
think of our predictive PCA as an unsupervised dimension reduction
approach. On the other hand, unlike traditional (sparse) PCA, the
approach is ``supervised'' by geographical data to increase
predictability of the PC scores. Alternatively, one can
potentially develop a dimension reduction approach that is fully
supervised by both geographical and health data. This, for
example, can be done using a Bayesian framework by introducing
lower dimensional latent variables which govern the joint
distribution between pollutants and the health variable. In joint
models, in other words, the first stage exposure model is a part
of a unified model for the exposure and health data (cf.
\cite{sinha2010semiparametric} in nutritional epidemiology and
\cite{gryparis2009measurement} in air pollution epidemiology). We
argue against such a joint approach for four reasons. First, using
a two-step approach in air pollution studies offers a way of
``cutting feedback" that may drastically influence estimation and
prediction in the exposure model in unexpected ways, especially
when the prediction model is misspecified. Second, when simpler
two-step approaches are used, performing a sensitivity analysis
using different exposure models in the first step can be done
easily to understand the effects of these models. Third, using two-step approaches allows use of the same predicted exposures
across studies with multiple health endpoints which makes such
approaches more practical and scientifically desirable. Fourth, an
inherent disadvantage of a unified approach is that it requires
specifying high dimensional joint distributions (and the
associated priors in a Bayesian framework). By simple virtue of
the high dimensionality, this task is computationally much more
expensive than a two-stage approach, particular in the context of
air pollution studies where there are less than 1000 monitors that
are used in the prediction step and a number of subjects that are
in the 10s of thousands in the health analysis step. For more
details on discussion between unified and two-step approaches,
readers are referred to \cite{szpiro2013measurement}.

In the literature, when both response variables (as our
multi-pollutant data) and covariates (as our GIS covariates and
thin-plate splines) are multivariate, methods known as two-block
partial least-squares (two-block PLS) can be used for analyzing the covariance
between the two sets of variables
\citep[cf.][]{sampson1989neurobehavioral, streissguth1993enduring,
rohlf2000use}. These methods are similar to the well-known method
of canonical correlation analysis \citep{jackson2005user}. Even
though two-block PLS is related to our PCA approach, in predictive
PCA, our goal is quite distinct: we want to improve predictability
of the PC scores using the same GIS covariates and splines and to
explain most of the variability in the multi-pollutant data. In
contrast to our goal, in two-block PLS and canonical correlation
analysis, the two sets of variables are treated symmetrically in
an attempt to find possible relationships between the underlying
latent variables derived from the sets. These approaches are
useful for exploring the combinations of variables in the two sets
that account most for whatever covariation there is between them.

In this paper, we considered predictive PCA both with and without
a lasso-type penalty on the loading vectors to induce sparseness.
In principle, we could also add additional penalty parameter to
the $\alpha$ coefficients in Equation (\ref{eq:newprob}), which
could be particularly useful for incorporating richer spatial
prediction models by allowing for a way to eliminate insignificant
geographical covariates. This may, however, substantially slow
down our iterative optimization algorithm described in Section
\ref{subsec:algorithm}. As future research, a related idea is to
exploit the link between random effects and penalization
\citep[cf.][]{ruppert2003semiparametric, verbyla1999analysis} to
develop an algorithm that uses a random effects model for the
$\alpha$. Another promising venue for future research is to
develop a predictive dimension reduction approach that finds all
principal loadings and scores in one step. In contrast to our
sequential approach, this would facilitate borrowing information
between scores in order to potentially further improve
predictability.

{When variability of high dimensional and correlated variables is of interest in the context of reducing the dimensionality of the data, factor analysis is another widely used statistical technique \citep[cf.][]{harman1960modern}. In factor analysis, the original variables are defined as linear combinations of the unobserved variables called factors with error terms using a statistical model, whereas in PCA, the components scores are calculated as linear combinations of the original variables. Therefore, the main goal in factor analysis is to explain the correlations between the variables with these lower dimensional factors to understand what constructs underlie the data. In PCA, our goal is to extract the lower dimensional scores to reduce the dimensionality of the correlated observed variables to a smaller set of important independent composite variables in an interpretable way for subsequent use in the health analysis.

In our approach, we achieve sparsity by adding a penalty term to the optimization problem. This allows for an easy way to incorporate the GIS covariates into a new optimization problem in the predictive PCA. In both PCA and factor analysis, in order to make the interpretation of the scores or factors easier, one can also achieve sparsity via rotation of the loadings. There are two main types of rotations that can be used: orthogonal, where orthogonality of the loadings is preserved and oblique, when the new loadings are not necessarily required to be orthogonal to each other. The most commonly used technique based on rotation of the loadings is called varimax rotation \citep[cf.][]{kaiser1958varimax}. Similar to sparse PCA, varimax rotation can result in scores or factors that have a small number of large loadings and a large number of zero loadings. Potentially, we note that it may also be possible to develop an alternative predictive sparse PCA approach based on rotation techniques.}

\section*{Acknowledgements}

This manuscript was made possible by USEPA grant (RD-83479601-0)
and NIH/NIEHS grant (R01-ES020871-03). Its contents are solely the
responsibility of the grantee and do not necessarily represent the
official views of the USEPA or NIH/NIEHS. Further, USEPA does not
endorse the purchase of any commercial products or services
mentioned in the publication.

\bibliographystyle{apalike}
\bibliography{pca.papers}

\begin{thebibliography}{}

\bibitem[Abdi, 2003]{abdi2003partial}
Abdi, H. (2003).
\newblock Partial least squares regression (pls-regression).
\newblock {\em Encyclopedia for research methods for the social sciences},
  pages 792--795.

\bibitem[Analitis et~al., 2014]{analitis2014effects}
Analitis, A., Michelozzi, P., D’Ippoliti, D., de’Donato, F., Menne, B.,
  Matthies, F., Atkinson, R.~W., I{\~n}iguez, C., Basaga{\~n}a, X., Schneider,
  A., et~al. (2014).
\newblock Effects of heat waves on mortality: effect modification and
  confounding by air pollutants.
\newblock {\em Epidemiology}, 25(1):15--22.

\bibitem[Anderson, 2003]{and:ims:2003}
Anderson, T.~W. (2003).
\newblock {\em {An Introduction to Multivariate Statistical Analysis.}}
\newblock Wiley Series in Probability and Statistics.

\bibitem[Bell and Davis, 2001]{bell2001reassessment}
Bell, M.~L. and Davis, D.~L. (2001).
\newblock {Reassessment of the lethal London fog of 1952: novel indicators of
  acute and chronic consequences of acute exposure to air pollution.}
\newblock {\em Environmental health perspectives}, 109(Suppl 3):389.

\bibitem[Bergen et~al., 2013]{bergen2013national}
Bergen, S., Sheppard, L., Sampson, P.~D., Kim, S.-Y., Richards, M., Vedal, S.,
  Kaufman, J.~D., and Szpiro, A.~A. (2013).
\newblock A national prediction model for pm2.5 component exposures and
  measurement error--corrected health effect inference.
\newblock {\em Environmental health perspectives}, 121(9):1017.

\bibitem[Brauer et~al., 2003]{brauer2003estimating}
Brauer, M., Hoek, G., van Vliet, P., Meliefste, K., Fischer, P., Gehring, U.,
  Heinrich, J., Cyrys, J., Bellander, T., Lewne, M., et~al. (2003).
\newblock Estimating long-term average particulate air pollution
  concentrations: application of traffic indicators and geographic information
  systems.
\newblock {\em Epidemiology}, 14(2):228--239.

\bibitem[Brook et~al., 2007]{brook2007further}
Brook, J.~R., Burnett, R.~T., Dann, T.~F., Cakmak, S., Goldberg, M.~S., Fan,
  X., and Wheeler, A.~J. (2007).
\newblock {Further interpretation of the acute effect of nitrogen dioxide
  observed in Canadian time-series studies}.
\newblock {\em Journal of Exposure Science and Environmental Epidemiology},
  17:S36--S44.

\bibitem[Bryant and Yarnold, 1995]{bryant1995principal}
Bryant, F.~B. and Yarnold, P.~R. (1995).
\newblock {Principal components analysis and exploratory and confirmatory
  factor analysis. In L. G. Grimm \& P. R. Yarnold (Eds.).}
\newblock {\em Reading and understanding multivariate statistics}, pages
  99--136.

\bibitem[Cattell, 1966]{cattell1966scree}
Cattell, R.~B. (1966).
\newblock The scree test for the number of factors.
\newblock {\em Multivariate behavioral research}, 1(2):245--276.

\bibitem[Crouse et~al., 2010]{crouse2010postmenopausal}
Crouse, D.~L., Goldberg, M.~S., Ross, N.~A., Chen, H., and Labr{\`e}che, F.
  (2010).
\newblock {Postmenopausal breast cancer is associated with exposure to
  traffic-related air pollution in Montreal, Canada: a case--control study}.
\newblock {\em Environmental health perspectives}, 118(11):1578.

\bibitem[Delfino et~al., 2014]{delfino2014asthma}
Delfino, R.~J., Wu, J., Tjoa, T., Gullesserian, S.~K., Nickerson, B., and
  Gillen, D.~L. (2014).
\newblock Asthma morbidity and ambient air pollution: effect modification by
  residential traffic-related air pollution.
\newblock {\em Epidemiology}, 25(1):48--57.

\bibitem[Dubrule, 1984]{dubrule1984comparing}
Dubrule, O. (1984).
\newblock Comparing splines and kriging.
\newblock {\em Computers \& Geosciences}, 10(2):327--338.

\bibitem[Eldred et~al., 1988]{eldred1988improve}
Eldred, R.~A., Cahill, T.~A., and Pitchford, M. (1988).
\newblock {IMPROVE: A new remote area particulate monitoring system for
  visibility studies}.
\newblock {\em Proceedings of the 81st Annual Meeting of the Air Pollution
  Control Association, Dallas, TX.}

\bibitem[EPA, 2009]{epa2009integrated}
EPA (2009).
\newblock Integrated science assessment for particulate matter.
\newblock {\em US Environmental Protection Agency Washington, DC}.

\bibitem[Gorsuch, 1983]{Gorsuch1983principal}
Gorsuch, R.~L. (1983).
\newblock {\em Factor analysis (2nd ed.).}
\newblock Hillsdale, NJ: Lawrence Erlbaum Associates.

\bibitem[Gryparis et~al., 2009]{gryparis2009measurement}
Gryparis, A., Paciorek, C.~J., Zeka, A., Schwartz, J., and Coull, B.~A. (2009).
\newblock Measurement error caused by spatial misalignment in environmental
  epidemiology.
\newblock {\em Biostatistics}, 10(2):258--274.

\bibitem[Harman, 1967]{harman1960modern}
Harman, H.~H. (1967).
\newblock {\em {Modern factor analysis, (2nd ed.)}}.
\newblock Univ. of Chicago Press.

\bibitem[Hoek et~al., 2008]{hoek2008review}
Hoek, G., Beelen, R., de~Hoogh, K., Vienneau, D., Gulliver, J., Fischer, P.,
  and Briggs, D. (2008).
\newblock A review of land-use regression models to assess spatial variation of
  outdoor air pollution.
\newblock {\em Atmospheric Environment}, 42(33):7561--7578.

\bibitem[Hutchinson and Gessler, 1994]{hutchinson1994splines}
Hutchinson, M. and Gessler, P. (1994).
\newblock Splines—more than just a smooth interpolator.
\newblock {\em Geoderma}, 62(1):45--67.

\bibitem[Jackson, 2005]{jackson2005user}
Jackson, J.~E. (2005).
\newblock {\em A user's guide to principal components}, volume 587.
\newblock John Wiley \& Sons.

\bibitem[Jerrett et~al., 2005]{jerrett2005spatial}
Jerrett, M., Burnett, R.~T., Ma, R., Pope~III, C.~A., Krewski, D., Newbold,
  K.~B., Thurston, G., Shi, Y., Finkelstein, N., Calle, E.~E., et~al. (2005).
\newblock Spatial analysis of air pollution and mortality in {Los Angeles}.
\newblock {\em Epidemiology}, 16(6):727--736.

\bibitem[Jolliffe, 1986]{jolliffe1986principal}
Jolliffe, I.~T. (1986).
\newblock {\em Principal component analysis}, volume 487.
\newblock Springer-Verlag New York.

\bibitem[Kaiser, 1958]{kaiser1958varimax}
Kaiser, H.~F. (1958).
\newblock The varimax criterion for analytic rotation in factor analysis.
\newblock {\em Psychometrika}, 23(3):187--200.

\bibitem[Kaiser, 1960]{kaiser1960application}
Kaiser, H.~F. (1960).
\newblock The application of electronic computers to factor analysis.
\newblock {\em Educational and psychological measurement}.

\bibitem[Kim and Mueller, 1978]{kim1978factor}
Kim, J.-O. and Mueller, C.~W. (1978).
\newblock {\em Factor analysis: Statistical methods and practical issues},
  volume~14.
\newblock Sage.

\bibitem[Kim et~al., 2009]{kim2009health}
Kim, S.-Y., Sheppard, L., and Kim, H. (2009).
\newblock Health effects of long-term air pollution: influence of exposure
  prediction methods.
\newblock {\em Epidemiology}, 20(3):442--450.

\bibitem[K{\"u}nzli et~al., 2005]{kunzli2005ambient}
K{\"u}nzli, N., Jerrett, M., Mack, W.~J., Beckerman, B., LaBree, L., Gilliland,
  F., Thomas, D., Peters, J., and Hodis, H.~N. (2005).
\newblock Ambient air pollution and atherosclerosis in {Los Angeles}.
\newblock {\em Environmental health perspectives}, pages 201--206.

\bibitem[Lee et~al., 2015]{lee2015impact}
Lee, A., Szpiro, A., Kim, S., and Sheppard, L. (2015).
\newblock Impact of preferential sampling on exposure prediction and health
  effect inference in the context of air pollution epidemiology.
\newblock {\em Environmetrics}, 26(4):255--267.

\bibitem[Logan, 1953]{logan1953mortality}
Logan, W. (1953).
\newblock {Mortality in the London fog incident, 1952}.
\newblock {\em The Lancet}, 261(6755):336--338.

\bibitem[Matheron, 1981]{matheron1981splines}
Matheron, G. (1981).
\newblock Splines and kriging: their formal equivalence.
\newblock {\em Down-to-earth statistics: solutions looking for geological
  problems}, 8:77--95.

\bibitem[Mercer et~al., 2011]{mercer2011comparing}
Mercer, L.~D., Szpiro, A.~A., Sheppard, L., Lindstr{\"o}m, J., Adar, S.~D.,
  Allen, R.~W., Avol, E.~L., Oron, A.~P., Larson, T., Liu, L.-J.~S., et~al.
  (2011).
\newblock Comparing universal kriging and land-use regression for predicting
  concentrations of gaseous oxides of nitrogen for the multi-ethnic study of
  atherosclerosis and air pollution (mesa air).
\newblock {\em Atmospheric Environment}, 45(26):4412--4420.

\bibitem[Miller, 1998]{miller1998handbook}
Miller, G.~J. (1998).
\newblock {\em Handbook of research methods in public administration}, volume
  134.
\newblock CRC press.

\bibitem[Miller et~al., 2007]{miller2007long}
Miller, K.~A., Siscovick, D.~S., Sheppard, L., Shepherd, K., Sullivan, J.~H.,
  Anderson, G.~L., and Kaufman, J.~D. (2007).
\newblock Long-term exposure to air pollution and incidence of cardiovascular
  events in women.
\newblock {\em New England Journal of Medicine}, 356(5):447--458.

\bibitem[Nemery et~al., 2001]{nemery2001meuse}
Nemery, B., Hoet, P.~H., and Nemmar, A. (2001).
\newblock {The Meuse Valley fog of 1930: an air pollution disaster}.
\newblock {\em The Lancet}, 357(9257):704--708.

\bibitem[Park et~al., 2008]{park2008air}
Park, S.~K., O’Neill, M.~S., Vokonas, P.~S., Sparrow, D., Wright, R.~O.,
  Coull, B., Nie, H., Hu, H., and Schwartz, J. (2008).
\newblock Air pollution and heart rate variability: effect modification by
  chronic lead exposure.
\newblock {\em Epidemiology (Cambridge, Mass.)}, 19(1):111.

\bibitem[Peres-Neto et~al., 2005]{peres2005many}
Peres-Neto, P.~R., Jackson, D.~A., and Somers, K.~M. (2005).
\newblock How many principal components? stopping rules for determining the
  number of non-trivial axes revisited.
\newblock {\em Computational Statistics \& Data Analysis}, 49(4):974--997.

\bibitem[Pope~III et~al., 2002]{pope2002lung}
Pope~III, C.~A., Burnett, R.~T., Thun, M.~J., Calle, E.~E., Krewski, D., Ito,
  K., and Thurston, G.~D. (2002).
\newblock Lung cancer, cardiopulmonary mortality, and long-term exposure to
  fine particulate air pollution.
\newblock {\em Jama}, 287(9):1132--1141.

\bibitem[Pope~III and Dockery, 2006]{pope2006health}
Pope~III, C.~A. and Dockery, D.~W. (2006).
\newblock Health effects of fine particulate air pollution: lines that connect.
\newblock {\em Journal of the Air \& Waste Management Association},
  56(6):709--742.

\bibitem[Rohlf and Corti, 2000]{rohlf2000use}
Rohlf, F.~J. and Corti, M. (2000).
\newblock Use of two-block partial least-squares to study covariation in shape.
\newblock {\em Systematic Biology}, 49(4):740--753.

\bibitem[Ruppert et~al., 2003]{ruppert2003semiparametric}
Ruppert, D., Wand, M.~P., and Carroll, R.~J. (2003).
\newblock {\em Semiparametric regression}.
\newblock Cambridge University Press.

\bibitem[Samet et~al., 2000]{samet2000fine}
Samet, J.~M., Dominici, F., Curriero, F.~C., Coursac, I., and Zeger, S.~L.
  (2000).
\newblock {Fine particulate air pollution and mortality in 20 U.S. cities,
  1987-1994}.
\newblock {\em New England journal of medicine}, 343(24):1742--1749.

\bibitem[Sampson et~al., 2013]{sampson2013regionalized}
Sampson, P.~D., Richards, M., Szpiro, A.~A., Bergen, S., Sheppard, L., Larson,
  T.~V., and Kaufman, J.~D. (2013).
\newblock A regionalized national universal kriging model using partial least
  squares regression for estimating annual pm 2.5 concentrations in
  epidemiology.
\newblock {\em Atmospheric Environment}, 75:383--392.

\bibitem[Sampson et~al., 1989]{sampson1989neurobehavioral}
Sampson, P.~D., Streissguth, A.~P., Barr, H.~M., and Bookstein, F.~L. (1989).
\newblock Neurobehavioral effects of prenatal alcohol: Part ii. partial least
  squares analysis.
\newblock {\em Neurotoxicology and teratology}, 11(5):477--491.

\bibitem[Sampson et~al., 2011]{sampson2011pragmatic}
Sampson, P.~D., Szpiro, A.~A., Sheppard, L., Lindstr{\"o}m, J., and Kaufman,
  J.~D. (2011).
\newblock Pragmatic estimation of a spatio-temporal air quality model with
  irregular monitoring data.
\newblock {\em Atmospheric Environment}, 45(36):6593--6606.

\bibitem[Shen and Huang, 2008]{shen2008sparse}
Shen, H. and Huang, J.~Z. (2008).
\newblock Sparse principal component analysis via regularized low rank matrix
  approximation.
\newblock {\em Journal of multivariate analysis}, 99(6):1015--1034.

\bibitem[Sinha et~al., 2010]{sinha2010semiparametric}
Sinha, S., Mallick, B.~K., Kipnis, V., and Carroll, R.~J. (2010).
\newblock {Semiparametric Bayesian analysis of nutritional epidemiology data in
  the presence of measurement error}.
\newblock {\em Biometrics}, 66(2):444--454.

\bibitem[Streissguth et~al., 1993]{streissguth1993enduring}
Streissguth, A.~P., Bookstein, F.~L., Sampson, P.~D., and Barr, H.~M. (1993).
\newblock {\em The enduring effects of prenatal alcohol exposure on child
  development: Birth through seven years, a partial least squares solution.}
\newblock The University of Michigan Press.

\bibitem[Szpiro and Paciorek, 2013]{szpiro2013measurement}
Szpiro, A.~A. and Paciorek, C.~J. (2013).
\newblock Measurement error in two-stage analyses, with application to air
  pollution epidemiology.
\newblock {\em Environmetrics}, 24(8):501--517.

\bibitem[Tibshirani, 1996]{tibshirani1996regression}
Tibshirani, R. (1996).
\newblock Regression shrinkage and selection via the lasso.
\newblock {\em Journal of the Royal Statistical Society. Series B
  (Methodological)}, 58:267--288.

\bibitem[Vedal et~al., 2012]{vedal2012pm}
Vedal, S., Kaufman, J., Larson, T., Sampson, P., Sheppard, L., Simpson, C.,
  Szpiro, A., McDonald, J., Lund, A., and Campen, M. (2012).
\newblock {University of Washington/Lovelace Respiratory Research Institute
  National Particle Component Toxicity (NPACT) Initiative}: Integrated
  epidemiological and toxicological cardiovascular studies to identify toxic
  components and sources of fine particulate matter (draft).
\newblock {\em Heath Effects Institute, Boston, MA, 2012}.

\bibitem[Verbyla et~al., 1999]{verbyla1999analysis}
Verbyla, A.~P., Cullis, B.~R., Kenward, M.~G., and Welham, S.~J. (1999).
\newblock The analysis of designed experiments and longitudinal data by using
  smoothing splines.
\newblock {\em Journal of the Royal Statistical Society: Series C (Applied
  Statistics)}, 48(3):269--311.

\bibitem[Wood, 2003]{wood2003thin}
Wood, S.~N. (2003).
\newblock Thin plate regression splines.
\newblock {\em Journal of the Royal Statistical Society: Series B (Statistical
  Methodology)}, 65(1):95--114.

\end{thebibliography}

\end{document}